\documentclass[11pt, letterpaper]{article}

\usepackage{amsmath, amsthm, amssymb}
\usepackage{bm, bbm}
\usepackage[ruled, linesnumbered]{algorithm2e}
\usepackage[utf8]{inputenc}
\usepackage[twoside]{geometry}
\usepackage{blindtext}
\usepackage{hyperref}
\usepackage{rotating, lscape}
\usepackage[toc,page]{appendix}
\usepackage{placeins}
\usepackage{array}
\usepackage{multicol}
\usepackage{sectsty}
\usepackage[english]{babel}

\usepackage[round]{natbib}

\usepackage{multirow}
\usepackage{booktabs}
\usepackage{array}
\usepackage{enumitem}
\usepackage{ragged2e}
\usepackage{longtable}
\usepackage{tikz}
\usepackage{pgfplots}
\usepackage{graphicx}
\usepackage{here}
\usepackage{makecell}
\usepackage{setspace}
\usepackage{colortbl}
\usepackage[justification=justified]{caption}
\pgfplotsset{compat=newest}

\usepackage{datetime}

\newdateformat{dmyDate}{\twodigit{\THEDAY}. \monthname[\THEMONTH] \THEYEAR}
\newdateformat{mdyDate}{\monthname[\THEMONTH] \THEDAY, \THEYEAR}

\definecolor{myDarkBlue}{RGB}{0,0,255}
\definecolor{g0}{RGB}{255,255,255}
\definecolor{g1}{RGB}{230,230,230}
\definecolor{g2}{RGB}{206,206,206}
\definecolor{g3}{RGB}{182,182,182}
\definecolor{g4}{RGB}{158,158,158}

\newcommand{\Keywords}[1]{\par\noindent{\small{\em Keywords\/}: #1}}
\newcommand{\hl}[1]{\textcolor{myDarkBlue}{#1}}

\let\Oldsection\section
\renewcommand{\section}{\FloatBarrier\Oldsection}

\DeclareMathOperator*{\argmax}{arg\,max}
\newcommand*{\defeq}{\mathrel{\vcenter{\baselineskip0.5ex \lineskiplimit0pt \hbox{\scriptsize.}\hbox{\scriptsize.}}} =}

\newcommand{\tablistcommand}{
  \leavevmode\par\vspace{-\baselineskip}
}
\newlist{tabitemize}{itemize}{1}
\setlist[tabitemize]{
  leftmargin=*,
  label=\textbullet,
  nosep,
  before=\tablistcommand,
  after=\tablistcommand
}

\newcolumntype{L}[1]{>{\raggedright\let\newline\\\arraybackslash\hspace{0pt}}m{#1}}
\newcolumntype{C}[1]{>{\centering\let\newline\\\arraybackslash\hspace{0pt}}m{#1}}
\newcolumntype{R}[1]{>{\raggedleft\let\newline\\\arraybackslash\hspace{0pt}}m{#1}}

\renewcommand{\bibpreamble}{\begin{multicols}{2}}
\renewcommand{\bibpostamble}{\end{multicols}}

\addto\extrasenglish{

}

\renewenvironment{abstract}
 {\small
  \begin{center}
  \bfseries \abstractname\vspace{-.5em}\vspace{0pt}
  \end{center}
  \list{}{
    \setlength{\leftmargin}{.5cm}%
    \setlength{\rightmargin}{\leftmargin}%
  }%
  \item\relax}
 {\endlist}

\begin{document}

\pagenumbering{Alph}

\begin{titlepage}

    \title{{\large \bf Static and Dynamic Models for Multivariate Distribution Forecasts:\\ Proper Scoring Rule Tests of  Factor-Quantile vs. Multivariate GARCH Models}}

        \author{\large Carol Alexander, Yang Han and Xiaochun Meng\\University of Sussex Business School}

    \date{10 December 2021}
    \maketitle
    \vspace{-1cm}
    \onehalfspacing

    \begin{abstract}
        \normalsize
        \noindent A plethora of static and dynamic models exist to forecast Value-at-Risk and other quantile-related metrics used in financial risk management. Industry practice tends to favour simpler, static models such as historical simulation or its variants whereas most academic research centres on dynamic models in the GARCH family. While numerous studies examine the accuracy of multivariate models for forecasting risk metrics, there is little research on accurately predicting the entire multivariate distribution. Yet this is an essential element of asset pricing or portfolio optimization problems having non-analytic solutions. We approach this highly complex problem using a variety of proper multivariate scoring rules to evaluate forecasts of eight-dimensional multivariate distributions: of exchange rates, interest rates and commodity futures. This way we test the performance of static models, viz. empirical distribution functions and a new factor-quantile model, with commonly used dynamic models in the asymmetric multivariate GARCH class. 
    \end{abstract}
    
    \bigskip
     
    \singlespacing
    \Keywords{Bagging; Continuous Ranked Probability Score; Energy Score; Factor Quantile Regression; Historical Simulation; Multivariate Density; forecast; Variogram Score}
    \thispagestyle{empty}

\end{titlepage}

\pagenumbering{arabic}
\renewcommand{\thefootnote}{\arabic{footnote}}
\doublespacing

\section{Introduction}
\label{Section:Introduction}
A plethora of static and dynamic models are employed to forecast Value-at-Risk (VaR) and other quantile-related metrics used in financial risk management. Many surveys of this vast literature have been published during the last twenty years, most recently by \cite{Nieto2016}, who provide a comprehensive review of the main methodological and empirical developments in (univariate) VaR models and their backtesting.
By contrast, developing tractable models for forecasting the entire distribution has attracted little academic attention, especially in a multivariate setting. The problem is very important nevertheless, because accurate  distribution forecasting is fundamental for the success of two important types of financial problems, viz. asset pricing, including the valuation of derivative products,\footnote{See for instance, \cite{Semenov2008}, \cite{Chiang2019} and \cite{Zhou2019}.} and the optimization of portfolio allocations when the decision-maker’s utility function and/or the distribution of the asset returns preclude the existence of an analytic solution.\footnote{In portfolio optimization a forecast of the entire multivariate distribution for asset returns is required to calculate the investor's expected utility -- see \cite{Ebens2009}, \cite{Lwin2017},  \cite{Thomann2020} and \cite{Grant2020}. There are many other studies, of course, but we have selected these to make the point that static models based on historical simulation are commonly employed by asset managers, because they are much simpler than dynamic models. Also, the academic literature in this area tends to focus more on modelling the decision-maker's utility than  on the underlying multivariate returns process. See \citet{birge2007} and \citet{resta2012} for reviews. }

A very popular topic for academic research is testing the accuracy of quantile forecasts from the \citet{bollerslev1986} Generalised Autoregressive Conditional Heteroscedasticity (GARCH) model -- and its numerous variants (see, for example, \citealt{kuester2006value} and \citealt{orhan2012comparison}). Yet, there is little evidence for the widespread adoption of such models by the industry, particularly in a large-scale multivariate setting.  Indeed, when commercial banks and other financial institutions report market risks they tend to favour simple static models such as historical simulation.\footnote{See \cite{Pritsker2006}, \cite{Berkowitz2011}, \cite{Prorokowski2014} \cite{Scheller2018} and many others. A survey by \cite{Perignon2010}  reported that almost 75\% of banks in their sample forecasted VaR using historical simulation.} The popularity of this approach to one-step-ahead forecasts is also supported by its robustness.\footnote{\cite{Cont2010} introduce a rigorous framework for studying this feature, showing that historical VaR is more robust than sophisticated risk metrics based on parametric models estimated by maximum likelihood.} More recently \cite{Danielsson2016} show that -- for predicting quantiles -- it remains unclear whether a complex dynamic model in the GARCH class outperforms one that is based on the idea that the next period joint distribution of the variables can be well approximated by their joint historical distribution, as for instance in \cite{Semenov2008}.  
Nevertheless, most academic \textcolor{blue}{research on the  forecasting accuracy of quantile-based risk metrics centres} on dynamic models in the GARCH family. 

This paper examines the accuracy of simple, static models that are typically favoured by financial institutions for predicting -- not just quantile -- but an entire multivariate distribution. Can static models produce more accurate forecasts than the complex dynamic models receiving the most attention in the academic literature? To answer this question  we consider a semi-parametric extension of historical simulation that generates a multivariate distribution using a parametric copula with empirical distribution function (EDF) marginals \citep{patton2009}. In addition, a substantial methodological section of this paper introduces a semi-parametric model for estimating multivariate distributions where marginals are derived from factor model quantile regressions \citep{koenker1982} and the dependence structure is modelled using a conditional copula \citep{patton2006}. We call it the \textit{Factor Quantile} (FQ) model. Both these static models are relative easy for less-quantitative mangers to comprehend, they scale naturally to very large dimensions and calibration is extremely fast. Because they are simpler, much quicker and more robust than multivariate GARCH models, these static models would be a natural candidate for adoption by the industry, provided we can show that they produce forecasts that are at least as accurate as the most common multivariate GARCH models.

To this end, we report a very comprehensive study that is the first extensive empirical evaluation of forecasting accuracy using the model confidence set approach of \cite{hansen2011} based on several proper multivariate scoring rules. Previously developed in meteorology and other branches of atmospheric science \citep{jolliffe2003, keune2014}, we apply these rules to assess the accuracy of daily forecasts for three different financial systems: exchange rates, interest rates and commodity futures. First we test univariate distribution forecasts using the weighted conditional ranked probability score proposed by \citet{gneiting2011}, which has the advantage of allowing different weight functions to focus on specific parts of the distribution. Then we apply the energy and variogram scores to measure the accuracy of multivariate distribution forecasts -- see \citet{gneiting2008b} and \citet{scheuerer2015}. In each case we assess the relative accuracy of the entire set of distribution forecasts considered in our empirical study through the equivalence test and elimination rules of the model confidence set of \citet{hansen2011}. 

This way we compare the forecasting performance of EDF models with two latent factor versions of the FQ model and with popular multivariate  GARCH models including the DCC-GARCH model of \cite{engle2002} with the exponential GARCH conditional variance specification of \cite{nelson1991} and Student-$t$ innovations. We use the Gaussian copula to reduce complexity and so that the EDF and FQ models scale naturally and easily to higher dimensions. However, the GARCH models are more difficult to scale and for this reason we consider eight-dimensional multivariate distributions, not higher dimensions. Our empirical study examines daily forecasts for eight USD exchange rates using data from 1999 -- 2018; a term structure of US interest rates with data from 1994 -- 2018; and eight  Bloomberg investable commodity indices from 1991 -- 2018.

In the following: \autoref{Section:Literature} sets our work in the context of the recent literature on multivariate distribution forecasts and on multivariate quantile models; \autoref{Section:FactorQuantileMethodology} introduces the general FQ methodology, illustrates this using a simple bivariate example, motivates various choices such as the quantile  grid and introduces two variants of the model with latent factors; \autoref{Section:EmpiricalStudy} presents our empirical study; \autoref{Section:Conclusions} summarises and concludes. All the code (in Python/MATLAB) and all three data sets used in this paper are available from the authors on request.

\section{Relevant Literature}\label{Section:Literature}

Forecasts of random variables should take the form of distributions to account for the randomness of the predicted event and prediction uncertainty. Yet most prior research in finance  limits empirical evaluation to point forecasts, often linked to quantiles, and typically of a univariate distribution. As a result, extensive empirical applications, even to univariate distribution forecasts, are hard to find -- possibly due to their computational complexity. Some applications of multivariate scoring rules can be found in the literature, but these are mostly limited to weather ensemble forecasts or they use the multivariate logarithmic score \citep{diks2010, diks2014} that has been criticised for its heavy penalty on low probability events which limits its practical application \citep{selten1998, gneiting2007b}. There are a few recent applications of proper scoring rules to financial or economic data, but these studies are far more limited than ours, and they are limited  to univariate distributions over a single out-of-sample period -- see \citet{panagiotelis2008}, \citet{ravazzolo2014} and \citet{alexander2019}.

 \citet{elliott2016} emphasise that a forecast is an economic decision which should be evaluated using a loss function. They survey some elementary tests for univariate forecasts based on loss differentials, such as \citet{diebold1995} who develop out-of-sample tests which compare errors of point forecasts, and \citet{giacomini2006} who extend these tests to multi-step point, interval or entire (univariate) distribution forecasts. 
 Since then several other papers investigate scoring rules applied to univariate forecasts: \citet{bao2007} advocate using the Kullback–Leibler information criterion which is derived from the logarithmic score; \citet{amisano2007} compare density forecasts using a weighted likelihood ratio test, but this is not a proper scoring rule; \citet{gneiting2007b} advocate using the Continuous Ranked Probability Score (CRPS); \citet{gneiting2011} extend the CRPS to adopt the weighting approach of \citet{amisano2007} so that evaluation can be focussed on a specific area of the distribution, such as a tail or the centre; and \citet{boero2011} find that ranked probability scores have better discriminatory power than logarithmic or quadratic scores. 
 
 Concerning the literature relevant to our proposed factor-quantile model, many empirical studies apply the quantile regression model of \citet{koenker1978} to predict financial data, but most examine the accuracy of a few specific (typically, extreme) quantiles and not the entire distribution. For instance, \citet{ma2008} and \citet{meligkotsidou2019} examine the predictability of stock returns and realized volatilities by lagged economic variables, and \citet{hua2013} use realized volatilities to predict quantiles of stock and bond returns comparing their quantile-regression results with four different univariate versions of GARCH. \citet{gaglianone2012} apply quantile regression to predict the distribution of U.S. unemployment rates, using a single-factor model with an exogenous consensus forecast based on forecast averaging. Similarly, \citet{bunn2016} use quantile factor models with exogenous forecasts of factors to predict the spot electricity price. \citet{cenesizoglu2008} use a model with a single lagged predictor variable to forecast quantiles and estimate the distribution by fitting a crude step function introduced by \citet{koenker1982}. Other papers such as \citet{hagforsetal2016} apply quantile regressions with multiple factors but use only in-sample diagnostics to examine model fit. Some papers use very short time-series -- e.g. \citet{koenker2010} -- and/or compare quantile regression with benchmark models which may be inadequate for the data. 

Concerning the various attempts to model multivariate quantiles, \citet{chakraborty2003} proposes to minimize a loss function that is a straightforward multivariate equivalent of the standard loss function used in univariate quantile regression, introduced by \citet{koenker1978}. However, this doesn't allow estimation of an associated distribution function because it is only based on the notion of geometric multivariate quantiles. Similarly, \citet{hallin2010} use the half-space depth contours of \citet{tukey1974} which are not equivalent to an associated distribution function. By contrast, insisting on the equivalence between the quantile function and a well-defined multivariate distribution, \citet{chavas2018} proposes that a multivariate $q$-quantile is a set $\mathbf{c}$ corresponding to the $q$-contour of the multivariate distribution $F$, i.e. $F(\mathbf{c}) = q$. This must reflect the general properties of $q$-quantiles, e.g. $F(\mathbf{c})$ is always non-decreasing -- however, the $q$-contours need not be convex and so $F$ need not have a unique inverse. \citet{chavas2018} assumes that quantiles are linear functions of exogenous variables. He only derives statistical properties of the quantile estimator when conditional distributions of the endogenous variables are independent. 


Several recent papers also examine new ways to predict the financial variables that we consider, albeit only by point forecasts. For the US interest rate term structure see \citet{almeida2017}; for USD exchange rates see \citet{greenaway2018}; and for commodity futures see \citet{zolotko2014} -- amongst many others. Finally, the voluminous literature on multivariate GARCH forecasting in financial markets is summarised by \citet{silvennoinen2009} and \citet{zakamulin2015}. Most studies only consider in-sample specification tests, with the exception of \citet{laurent2012}, who apply the model confidence set of \citet{hansen2011} and the \citet{hansen2005a} tests for superior predictive ability to GARCH covariance forecasts, not to multivariate returns distributions.


\section{Factor Quantile Models}
\label{Section:FactorQuantileMethodology}
Deriving multivariate distribution forecasts from a system of common factor quantile regressions presents many challenges. The basic problem is that there is no unique way to invert a multivariate distribution function and no inherent ordering of quantiles in multiple dimensions. So, unlike the univariate case, even the definition of a multivariate quantile is not unique and alternative definitions support different techniques for estimating multivariate quantile regressions, not all of which identify distribution functions. The motivation for FQ models is to circumvent these problems entirely, deriving a multivariate distribution by applying a conditional copula to marginals generated from univariate factor model quantile regressions. 

The fundamental steps are easily understood in three stages: (i) For each dependent variable, we predict a range of conditional \textcolor{blue}{quantile levels} in $(0,1)$ using univariate quantile regression on multiple common factors; (ii) For a given realisation of common factors, and for each dependent variable, fit a conditional distribution to the quantiles estimated in (i); and (iii) Impose a dependence structure on the conditional marginals using a conditional copula. 

This way, the FQ model generates a multivariate distribution, conditional on the common factors, with marginals derived from quantile regressions and with a flexible dependence structure imposed by the choice of copula. This algorithm is very fast and flexible, and because the quantile regressions are univariate it scales very easily as the dimension of the system increases. By comparison, multivariate quantile regression approaches, such as those proposed by \citet{chakraborty2003} or \citet{chavas2018}, require a vast data set, and are much more computationally intensive.

The starting point of our model description is a standard linear factor model
\begin{align*}
    \mathbf{y}_t=\bm{\alpha}+\mathbf{B}\mathbf{x}_t+\bm{\varepsilon}_t, \qquad \qquad t=1,\ldots, T,
\end{align*}
where: $\mathbf{y}_t=(\mathrm{y}_{1t},\ldots,\mathrm{y}_{nt})'$ and $\mathbf{x}_t=(\mathrm{x}_{1t},\ldots,\mathrm{x}_{mt})'$ denote the time $t$ values of $n$ dependent variables and $m$ common factors, respectively; $\bm{\alpha}=(\alpha_{1},\ldots,\alpha_{n})'$ is the vector of intercepts, and $\mathbf{B}$ is the matrix of factor sensitivities, both assumed constant; and $\bm{\varepsilon}_t=(\varepsilon_{1t},\ldots,\varepsilon_{nt})'$ is a multivariate error process.  Further, we assume that the data $\{\mathbf{y}_t\}^T_{t=1}$ are generated by a stochastic process $\mathbf{y}$ with stationary conditional joint distribution $F|\mathbf{x}$ and conditional marginal distributions $F_{1}|\mathbf{x},\ldots,F_{n}|\mathbf{x}$. 

Macroeconomic, fundamental and statistical factor models were introduced by \citet{ross1976}, \citet{fama1993} and \citet{connor2012} respectively. Applications to predicting stock portfolios, interest rates, exchange rates and economic variables have been considered by many authors, including \citet{patton2006}, \citet{coroneo2016}, \citet{duan2016}, 
and \citet{wellmann2018}. These apply standard estimation techniques, such as ordinary least squares, but then forecasts are limited to inferences on the means and variances of the dependent variables, conditional on each factor. By contrast, we use factor quantile regressions, which allow the explanatory variables to affect the dependent variables differently for each $\tau$-quantile, and estimation can trace out the conditional distribution of each dependent variable as $\tau$ ranges from $0$ to $1$. Thus, to capture this flexibility, we extend the contemporaneous quantile-regression framework of \citet{gaglianone2012} to multiple factors as follows: 
\begin{align}
    \label{Equation:QuantileModel}
    \mathbf{y}_t=\bm{\alpha}^{(\tau)}+\mathbf{B}^{(\tau)}\mathbf{x}_t+\bm{\varepsilon}_t^{(\tau)}, \qquad \qquad t=1,\ldots,T, 
\end{align}
where $\bm{\varepsilon}_{t}^{(\tau)}$ are quantile-dependent error processes, $\bm{\alpha}^{(\tau)}$ are the intercepts and $\mathbf{B}^{(\tau)}$ are the matrices of quantile regression coefficients. We can view $\mathbf{y}_t^{(\tau)}=\bm{\alpha}^{(\tau)}+\mathbf{B}^{(\tau)}\mathbf{x}_t$ as the vector containing the $\tau$-quantile of each element of $\mathbf{y}_t$, conditional on $\mathbf{x}_t$.

Motivated by the relatively weak fit of forecast models using lagged explanatory variables, especially when multiple quantiles are considered, we shall assume a contemporaneous relationship between dependent and explanatory variables. In the studies of \citet{cenesizoglu2008} and \citet{zhu2013}, most of the lagged economic predictors for the stock and bond returns are not statistically significant in the quantile regressions. By contrast, \citet{bunn2016} utilize contemporaneous information in their quantile model which performs well against asymmetric GARCH models with non-normal innovations. 

Thus, our starting point is similar to a multi-factor generalisation of the quantile regressions in \citet{gaglianone2012}. To derive forecasts for our dependent variables we shall need to set values for our independent variables. In general, FQ models may use any consistent set of values $\mathbf{x}^*$ which accounts for the dependency structure between the explanatory variables, and assuming such a set is available we can estimate the quantile regressions \eqref{Equation:QuantileModel} using historical data for $t=1,\ldots,T$, for some pre-defined set $\mathbbm{Q}$ of $\tau \in (0,1)$, and then predict each conditional quantile as: 
\begin{align}
    \label{Equation:MVQuantileModelEstimation}
    \mathbf{\hat{y}}^{(\tau)}|\mathbf{x}^* =\hat{\bm{\alpha}}^{(\tau)}+\hat{\mathbf{B}}^{(\tau)}\mathbf{x}^*. 
\end{align}
Next consider a quantile grid \textcolor{blue}{of quantile levels} $\mathbbm{Q}$ where $0<\tau<1$ for all $\tau\in \mathbbm{Q}$ and focus for now on the $i$-th element of $\mathbf{y}$. If $\mathbbm{Q}$ outlines a sufficiently dense grid, the shape of the entire conditional distribution function $F_{i}|\mathbf{x}^*$ of $\mathrm{y}_{i}$ can be estimated through $\{(\tau,\mathrm{\hat{y}}_{i}^{(\tau)}|\mathbf{x}^*): \tau\in \mathbbm{Q}\}$. The optimal node positions depend on $F_{i}|\mathbf{x}^*$ and should focus on parts where the distribution is expected to be irregular. Since fitting the tails of the distribution is more of a challenge than fitting the centre, nodes concentrated around the tails are beneficial. 

Multiple methods have been applied to interpolate a continuous distribution from the estimated quantiles: \citet{koenker1982} use a step function which assigns the value of the next smallest quantile in ${\tau\in \mathbbm{Q}}$. This method is adapted by \citet{cenesizoglu2008} and \citet{pedersen2015}; kernel density estimations, e.g. with Gaussian or Epanechnikov kernel, can be employed as in \citet{koenker2010} and \citet{gaglianone2012}; or shape-preservation can be maximized using the Piecewise Cubic Hermite Interpolating Polynomials (PCHIP) algorithm of \citet{fritsch1980}.   We explain our reasons for using the third alternative in the next section.

Given a vector ${\mathbf{x}}^*$ of values for the common factors, denote the interpolated conditional distribution functions by $\hat{F}_{i}|{\mathbf{x}}^*$, for $i=1,\ldots,n$. The probability integral transform variables are uniformly distributed if the forecast is probabilistically calibrated and will only be independent if the residuals $\varepsilon_{i}|{\mathbf{x}}^*={F}_{i}-\hat{F}_{i}|{\mathbf{x}}^*$ are independent which may be not the case unless the factor model perfectly represents the regressand without any missing variables or similar problems. Otherwise, we capture dependence using an extension of Sklar's theorem to conditional copulas due to \citet{patton2006} which represents a joint conditional distribution in terms of a unique conditional copula defined by
\begin{align}
    \label{Equation:Copula}
    \hat{F}\left(\mathbf{y}|\mathbf{x}^*\right)=C\left(\left.\hat{F}_{1}\left(y_1|\mathbf{x}^*\right),\ldots,\hat{F}_{n}\left(y_n|\mathbf{x}^*\right)\right|\mathbf{x}^*\right), 
\end{align}
where C denotes the conditional copula, which is a multivariate distribution function with marginal distributions that are uniform on [0, 1]. This way, any conditional marginals can be transformed into a valid multivariate distribution provided the copula is conditioned on the same variables as the marginal distributions. As \citet{patton2013} points out, this multi-stage approach results in a multivariate model without the challenges associated with simultaneous estimations in high dimensions. 

To summarize, the general methodology of FQ models proceeds as follows:
\begin{description}
    \item[Stage 1] Estimate quantile regressions for $\tau$-quantiles where $\tau\in (0,1)$ are pre-specified by a grid $\mathbbm{Q}$ \textcolor{blue}{of quantile levels}; 
    \item[Stage 2] For a given vector $\mathbf{x}^*$ for the common factors, interpolate over conditional quantiles in $\mathbbm{Q}$ to obtain each conditional marginal $\hat{F}_{1}|\mathbf{x}^*,\ldots,\hat{F}_{n}|\mathbf{x}^*$; 
    \item[Stage 3] Use a conditional copula and apply \eqref{Equation:Copula} to obtain the joint conditional distribution.
\end{description}

\subsection{A Simple Illustration of a Conditional Factor Quantile Model}
\label{Section:Example}
Consider the case where dependent variables are excess stock returns $r_{it}$ with $i=1,\ldots,n$ and the factor model is the two-factor Capital Asset Pricing Model (CAPM) introduced by \citet{kraus1976}. Through the inclusion of a quadratic term in the excess market return $r_{tM}$, the two-factor CAPM captures different sensitivities to positive and negative returns and allows the systematic risk of a stock to be related to skewness, as in \citet{harvey2000}. The quantile regressions are:
\begin{align}
    \label{Equation:QCAPM}
    r_{it}=\alpha^{(\tau)}+\beta^{(\tau)} r_{tM}+\gamma^{(\tau)} r_{tM}^2 + \varepsilon_{it}^{(\tau)}, \qquad t=1, \ldots, T \mbox{ and }  i=1, \ldots, n. 
\end{align}

All conditional quantiles of the quadratic CAPM in \eqref{Equation:QCAPM} are calibrated on data from 03 January 2000 to \textcolor{blue}{28 June 2018}. The market return is on the S\&P500 index and all distributions are conditional on the realized S\&P500 return on 29 June 2018. All data are of daily frequency, and we use 2000 observations to fit the quantile regression models.

\begin{figure}[!htb]
    \caption{Distribution estimates with varying quantile grids (Apple)}
    \vspace{-0.3cm}
     \caption*{\footnotesize {   Conditional distributions for the return on Apple based on an equidistant quantile grid with $|\mathbbm{Q}|=500$ (shaded area) are compared with distributions based on $|\mathbbm{Q}|=9$ (solid line). The step function and the shape-preserving interpolation utilize the smaller quantile grid with a focus on the tails while the kernel estimation uses equidistant nodes as illustrated with the rugs on the right-side axis since this yields better estimations. }}
    \label{Figure:DistributionComparison}
    \centering
    \includegraphics[width=\textwidth]{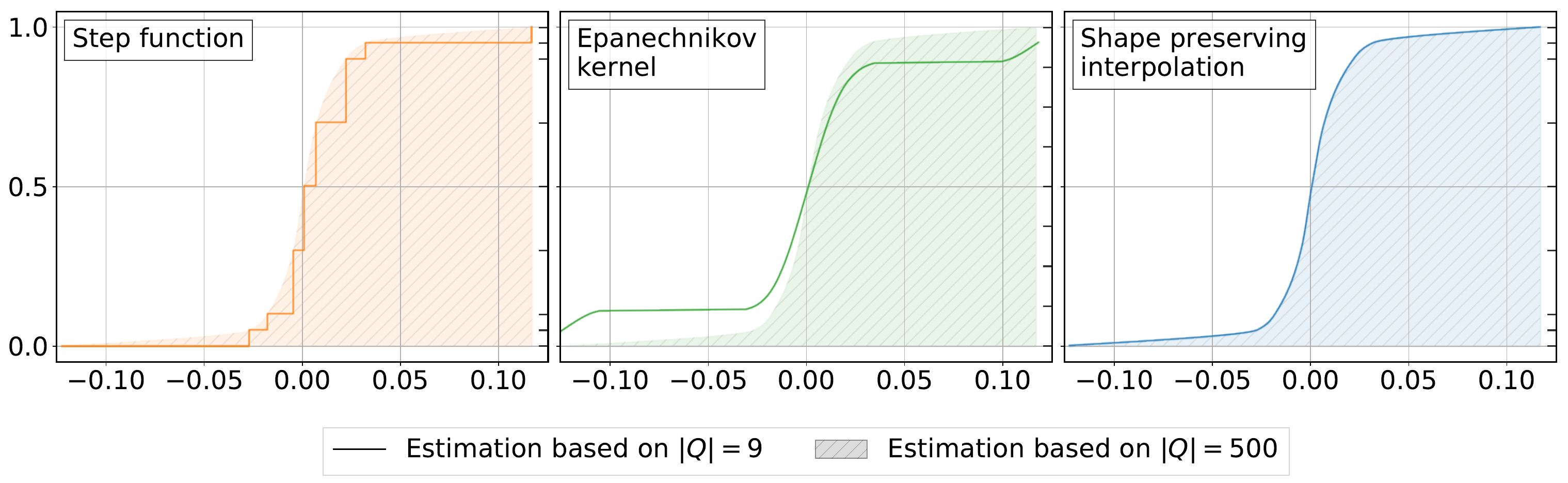}
\end{figure}

We start by illustrating the selection of the quantile grid, which itself depends on the choice of interpolation. To see this, let us compare the properties of three alternative interpolation methods reviewed in the previous section. We estimate quantile regressions for returns on the stock Apple with the S\&P500 as market factor and two different quantile grids \textcolor{blue}{with quantile levels} $\mathbbm{Q}$, one with $|\mathbbm{Q}|=9$ and another with $|\mathbbm{Q}|=500$. \textcolor{blue}{With $|\mathbbm{Q}|=500$ we use an equidistant quantile grid.} With $|\mathbbm{Q}|=9$ we use $\mathbbm{Q}=\{0.001, 0.05, 0.1, 0.3, 0.5, 0.7, 0.9, 0.95, 0.999\}$ which has more nodes in the extremes. This is to increase the domain of the estimated function without the need for extrapolation. Quantile regression is likely to yield high sampling error for these extreme nodes because there are fewer data points in those percentiles, by definition. But, on balance, taking account of the monotonicity requirement for quantiles and the hit-or-miss accuracy of ad-hoc extrapolation, additional nodes in the tails should benefit the accuracy of the estimated distribution. \autoref{Figure:DistributionComparison} compares the results for (i) the step function introduced by \citet{koenker1978} and applied by \citet{cenesizoglu2008}, on the left in orange; (ii) the Epanechnikov kernel advocated by \citet{gaglianone2012} in the middle in green; and (iii) the PCHIP interpolation, on the right in blue. 

\begin{table}[!htb]
    \centering
    \small
    \caption{Kolmogorov-Smirnov $p$-values of distribution comparison (Apple)}
     \vspace{-0.3cm}
    \caption*{\footnotesize {The quantiles for the return of Apple are calculated with the quadratic CAPM  \eqref{Equation:QCAPM} and
data from 03 January 2000 to 28 June 2018. We model the market return through the returns of
the S\&P500 index and condition all distributions on the realized S\&P return from 29 June 2018.}}
    \label{Table:ConvergenceDistribution}
    \begin{tabular}[!htb]{p{1cm}rr}
        \midrule
        \midrule
        $|\mathbbm{Q}|$ & Step function & Epanechnikov kernel \\ 
        \midrule
        10 & 0.0027 & 0.2562 \\
        20 & 0.4493 & 0.9154 \\
        30 & 0.8110 & 0.9855 \\
        40 & 0.9885 & 0.9996 \\
        50 & 0.9997 & 0.9996 \\
        \midrule
    \end{tabular}
\end{table}

The quantile grid with cardinality 500 produces very similar distributions for all three methods. These are indistinguishable in a Kolmogorov-Smirnov test at significance level of 1\%. However, with $|\mathbbm{Q}|=9$ the shape-preserving interpolation fits much better than the kernel or the step function, which yield vastly different distributions depending on the choice of $\mathbbm{Q}$. Only the shape-preserving interpolation produces similar results for both  grid sizes.  \cite{fritsch1980} show that the PCHIP method preserves monotonicity defined by the estimated quantiles so, provided the estimated quantiles do not cross, the PCHIP distribution will be well-defined.

The smaller the grid size, the  further apart the quantiles  and the less chance that estimated quantiles exhibit non-monotonic behaviour. To quantify the additional quantile  grid requirements of the kernel and the step function, we sample from distributions with varying equidistant quantile grids and compare them with the estimation based on $|\mathbbm{Q}|=500$ through a Kol\-mo\-go\-rov-Smirnov test. \autoref{Table:ConvergenceDistribution} lists the $p$-values. The kernel requires $|\mathbbm{Q}|=35$ and the step function $|\mathbbm{Q}|=50$ to achieve a similar distribution. \textcolor{blue}{Both  $|\mathbbm{Q}|=35$ and $|\mathbbm{Q}|=50$ use an equidistant quantile grid.} However, the shape-preserving interpolation with $|\mathbbm{Q}|=9$ yields a function which a Kolmogorov-Smirnov test cannot distinguish from the one based on $|\mathbbm{Q}|=500$ at a significance level of 1\%. The lower cardinality requirement of the shape-preserving interpolation is especially relevant in practice since it leads to major computational improvements. The total time taken for estimating all quantile regressions and then applying the distribution estimation with $|\mathbbm{Q}|=9$, $35$ and $50$, respectively, is over four times longer for both the kernel and the step function than the shape-preserving interpolation. 
Hence, in the rest of this paper we shall use the much faster and more accurate shape-preserving algorithm for interpolating all conditional distributions.

\begin{figure}[!htb]
    \caption{Conditional distribution and density forecasts (Apple and P\&G)}
    \vspace{-0.3cm}
        \caption*{\footnotesize The conditional marginal distribution and corresponding density for two US stock returns are generated with a FQ model based on the quadratic CAPM in \eqref{Equation:QCAPM}. For the calibration, we use the quantile grid with $|\mathbbm{Q}|=9$ as illustrated with the rugs on the right-side axis of the left figure.}
    \label{Figure:MarginalDistributionPlot}
    \centering
    \includegraphics[width=\textwidth]{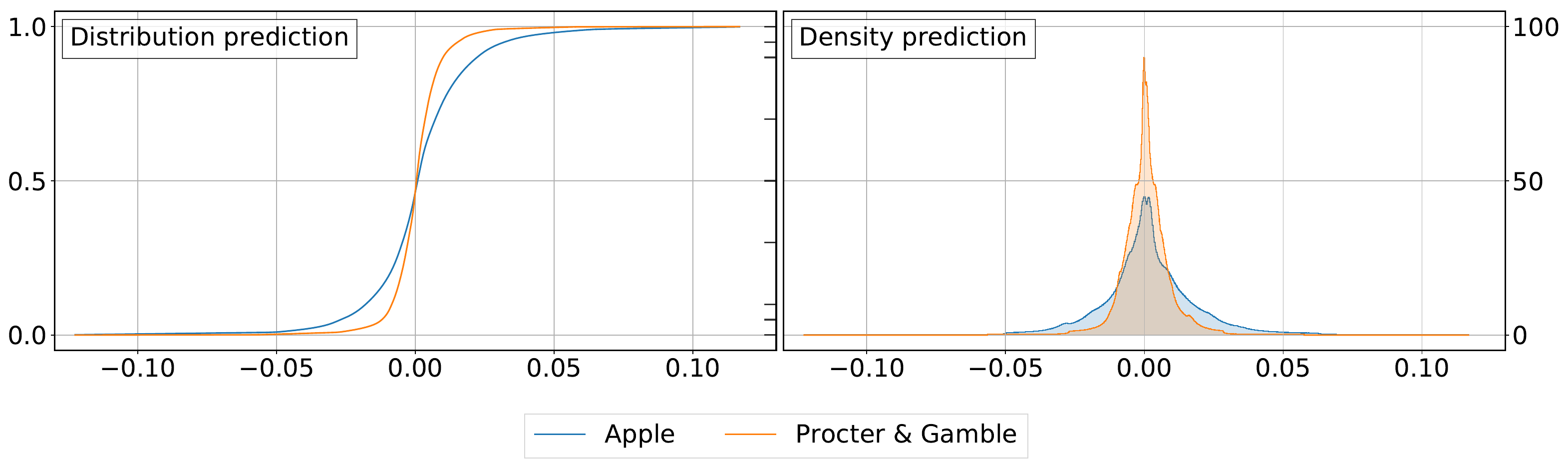}
\end{figure}

\begin{figure}[!htb]
    \caption{Density contours of the joint conditional density forecasts (Apple and P\&G)}
     \vspace{-0.3cm}
        \caption*{\footnotesize We use maximum likelihood estimation on the stock returns to derive the optimal parameters for the Gaussian and Archimedean copulas. This yields $\rho=0.1988$ for the Gaussian copula and $\theta=1.1590$ or $\theta=0.2690$ for the Gumbel and Clayton copula respectively. Apple returns are on the horizontal axis and PG returns are on the vertical axis.}
    \label{Figure:JointDistributionPlot}
    \centering
    \includegraphics[width=\textwidth]{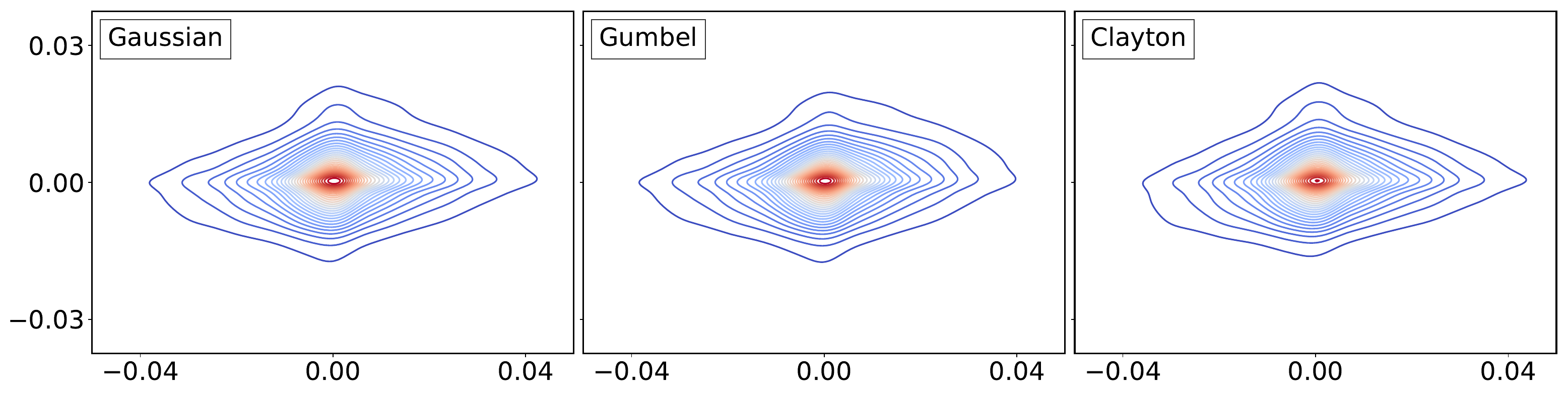}
\end{figure}

Next we estimate quantile regressions \eqref{Equation:QCAPM} for the quantile grid with $|\mathbbm{Q}|=9$ to another US stock, Procter and Gamble (P\&G) over the same time period. Interpolating allows for a visual comparison of the conditional distributions and densities of Apple and P\&G, depicted in \autoref{Figure:MarginalDistributionPlot}. Both distributions and densities exhibit irregularities which are difficult to capture with alternative parametric estimations. Now we use these conditional marginal distributions and fit conditional joint distributions with a Gaussian, a Gumbel and a Clayton copula. The density contours of the conditional joint densities are illustrated in \autoref{Figure:JointDistributionPlot}. These show slight but noticeable differences depending on the copula choice. \textcolor{blue}{Based on the standard information criteria, such as the Akaike Information Criterion (AIC) and the Bayesian Information Criterion (BIC), we prefer the Gumbel copula for this conditional joint distribution. It is also worth pointing out that the Archimedian copulae only have one parameter to model dependency structure, regardless of the number of dimensions of the multivariate distributions. Therefore, even though the Archimedian copulae can be easy to estimate, they inevitably become less and as less accurate as the portfolio size increases.}

We should emphasize that the dependency structure between the conditional marginal distributions of FQ models depends on the conditional copula. The quantile regression models \eqref{Equation:QuantileModel} share the same predictor variables $\mathbf{x}$, but this does not affect conditional rank correlation metrics such as Kendall's $\tau$ or Spearman's $\rho$. Of course, changes in values of $\mathbf{x}$ affect all dependent variables simultaneously, so the unconditional dependency depends on the common factor structure as well as the copula. 

\subsection{Static Latent Factor Quantile Models}
\label{Section:LatentFactors}
In this section we develop the FQ model when the common factors in equation \eqref{Equation:QuantileModel} are latent variables corresponding to the first principal components of the covariance matrix of $\mathbf{y}$. Following \citet{stock2002}, many papers on quantile regression employ principal components derived from the covariance matrix of a set of exogenous predictor variables. 
\citet{manzan2015} empirically evaluates the predictive power of principal components of a large number of exogenous macroeconomic indicators when used to augment the \citet{koenker2006} autoregressive model for quantiles. \citet{maciejowska2016} generalize the quantile regression averaging approach by \citet{nowotarski2015} with principal components to avoid the ex-ante model selection. Quantile regression averaging involves applying quantile regression with a set of individual point forecasts as independent variables and the observed value of the predicted variable as the dependent variable. By contrast, we are interested in the case that the latent factors are \textit{endogenous}, in the sense that the principal components are derived from the covariance matrix of the dependent variables alone. 
This endogenous approach was first employed by \citet{connor1993} who use asymptotic results on principal components to determine the appropriate number of factors for explaining returns on US stocks. The endogenous approach is desirable in practice, because it does not require extra data other than the dependent variables themselves.

In the following we consider a time-series sample $\mathbf{y}_t=(\mathrm{y}_{1t},\ldots,\mathrm{y}_{nt})'$ for $t = 1, \ldots, T$ on $n$ dependent variables. Latent factor quantile models are also applicable to cross-sectional data because the factors are entirely derived from the eigenvectors of a sample covariance matrix $\bm{\Sigma}$. Although the FQ approach could equally be applied to cross-sectional data, we assume a time-series setting, and we further assume $\mathbbm{E}(\mathbf{y}_t) = \mathbf{0}$ for all $t$ without loss of generality. This assumption is common for financial returns.

Denote the matrix with $j$-th column equal to the $j$-th eigenvector by $\mathbf{W}=(\mathbf{w}_1,\ldots,\mathbf{w}_n)$, having ordered these columns so that $\mathbf{w}_k$ is the unit eigenvector corresponding to $\lambda_k$, the $k$-th largest eigenvalue of $\bm{\Sigma}$. Set $\mathbf{p}_t=\mathbf{W}'\mathbf{y}_t=(\mathrm{p}_{1t},\ldots,\mathrm{p}_{nt})'$ so that $p_{kt}$ is the $k$-th principal component and its in-sample variance $\lambda_k$ decreases as $k$ increases. Because it is orthogonal, $\mathbf{W}'=\mathbf{W}^{-1}$, so inverting $\mathbf{W}$ yields the full principal component representation of the original system in terms of uncorrelated latent variables as $\mathbf{y}_t=\mathbf{W}\mathbf{p}_t$. Typically, the number of factors $m$ is selected so that a large fraction, but not all, of the total variance is explained, i.e. $m < n$.  Then a statistical factor model based on $m$ endogenous principal component factors is an approximate representation $\mathbf{y}_t \approx \mathbf{W}_m\mathbf{x}_{mt}$ where $\mathbf{W}_m$ denotes the first $m$ columns of $\mathbf{W}$ and $\mathbf{x}_{mt}=\left(\mathrm{p}_{1t},\ldots,\mathrm{p}_{mt}\right)'$. \textcolor{blue}{In the literature, it is often assumed that} a principal component representation based on the first $m$ components approximates the original data by omitting the variation captured by the last $n-m$ components, in effect disregarding  any importance this variation has for forecasting.

Now consider the choice of latent variables. There is a trade off between setting $m$ small enough to  ignore the variation that is regarded as unimportant and large enough to capture sufficient variation in the system to be informative for forecasts. Rules of thumb exist (such as taking $m$ to be large enough to capture at least 95\% of the variation with the other 5\% being assigned to information which is not useful for forecasts) but this is essentially a matter of empirical design which we discuss in more detail in \autoref{Section:Data}. Also, as in any factor model, forecasts for $\mathbf{y}_t$ are conditional on some predetermined values $\mathbf{x}^*_m$ for $\mathbf{x}_m$. Because our endogenous latent factors are contemporaneous, $\mathbf{x}_m$ is unknown. One approach is to fit a dynamic model for $\mathbf{x}_m$ to obtain an forecast $\mathbf{x}^*_m$, \textcolor{blue}{which can then be fed} into the factor quantile models. However, this approach defeats the simplicity of the factor quantile models. In this paper, we consider a static approach, where we use the unconditional distribution induced by the factor quantile models as our predictive density.

To obtain the static predictive density, we first consider the ``bootstrap aggregation'' -- commonly abbreviated to \textit{bagging} -- via an algorithm proposed by \citet{breiman1996}. When data $\mathbf{Z}$ are used in some model to obtain a \textcolor{blue}{distribution $\hat{F}$} the meta-algorithm generates $B$ bootstrap samples $\mathbf{Z}^{1}, \ldots, \mathbf{Z}^{B}$, each having the same pre-defined size, by drawing from $\mathbf{Z}$ with replacement. Then the density forecast based on bagging is the arithmetic average \textcolor{blue}{$\hat{F}^{\,\text{bag}}\defeq B^{-1}\sum^B_{b=1}\hat{F}^{b}$},
where \textcolor{blue}{$\hat{F}^{b}$} is the forecast based on data $\mathbf{Z}^{b}$. To proceed, we assume the quantile vector  $\hat{\mathbf{y}}^{(\tau)}$ follows a multivariate Gaussian distribution. The mean and variances of this multivariate Gaussian distribution can be computed as follows. Let us assume the parameter estimators are denoted as $\hat{\bm{\alpha}}^{(\tau)}_m$ and $\hat{\mathbf{B}}^{(\tau)}_m$, where \textcolor{blue}{$m$ indicates that these parameters are associated with $m$ principle components}.
The assumption $\mathbbm{E}(\mathbf{y}) = \mathbf{0}$,  together with our construction based on the principal component analysis,  implies that $\mathbb{E}\left(\mathbf{x}_{m}\right)=\mathbf{0}$. Hence, equation \eqref{Equation:MVQuantileModelEstimation} induces the following \textcolor{blue}{mean of each prediction quantile}:
\begin{align}
\label{Equation:Expectation}
\mathbb{E}\left(\hat{\mathbf{y}}^{(\tau)}\right)= \hat{\bm{\alpha}}^{(\tau)}_m. \end{align}
Further, since the principal components are uncorrelated with each other,  
the estimated conditional covariance between the $\tau_k$- and $\tau_l$-quantiles is:
\begin{align}
\label{Equation:Covariance} \mathbb{C}\mbox{ov}\left(\hat{\mathbf{y}}^{(\tau_{k})},\hat{\mathbf{y}}^{(\tau_{l})}\right)=\mathbb{C}\mbox{ov}\left(\hat{\bm{\alpha}}_{m}^{(\tau_{k})}+\hat{\mathbf{B}}_{m}^{(\tau_{k})}\mathbf{x}_{m}^{*},\hat{\bm{\alpha}}_{m}^{(\tau_{l})}+\hat{\mathbf{B}}_{m}^{(\tau_{l})}\mathbf{x}_{m}^{*}\right)\approx\hat{\mathbf{B}}_{m}^{(\tau_{k})}\mbox{diag}(\lambda_{1},\ldots,\lambda_{m})\hat{\mathbf{B}}_{m}^{(\tau_{l})'}.
\end{align}

\textcolor{blue}{Equations \eqref{Equation:Expectation} and \eqref{Equation:Covariance} can be used to define the multivariate Gaussian distribution for the quantile predictions of different levels for each variable, then we can obtain the bootstrap samples $\mathbf{Z}^{1}, \ldots, \mathbf{Z}^{B}$ from this multivariate Gaussian distribution}. 

 For each bagging draw, we  generate \textcolor{blue}{the distribution forecast $\hat{F}^b_{}$}, for $b=1,\ldots,B$. Then we average them, setting \textcolor{blue}{$$\hat{F}^{\,\text{bag}}_{}=B^{-1}\sum^B_{b=1}\hat{F}^b_{},$$} 
   Pseudocode for the bagging algorithm is \textcolor{blue}{shown in Algorithm \ref{Algorithm:FactorQuantileAB}}.

\begin{algorithm}
	\BlankLine
	\SetKwInOut{Input}{Input}
	\SetKwInOut{Output}{Output}
	\Input{Quantile grid of quantile levels $\mathbbm{Q}$ with $0<\tau<1$ for all \textcolor{blue}{$\tau_1,...,\tau_q \in \mathbbm{Q}$ with $q=|\mathbbm{Q}|$};\newline
		Observations on $\mathbf{y}_t$ for $t=1,\ldots,T$;
	}
	\Output{Unconditional multivariate distribution $\hat{F}_{T+1}$ of $\mathbf{y}_{T+1}$;}
	\BlankLine
	Use observations to calculate the first $m\leq n$ principal components $\mathbf{x}_t=(p_{1t},\ldots,p_{mt})$ where $m$ is determined by the target for the variance explained;
	
	\For{$i=1,\ldots,n$}{
		Estimate the factor quantile regressions 
		\begin{align*}
		\textcolor{blue}{\mathrm{y}_{it}\leftarrow\alpha_{im}^{(\tau_k)}+\mathbf{B}_{im}^{(\tau_k)}\mathbf{x}_t+\varepsilon^{(\tau_k)}_{it}}
		\end{align*}
		which yields \textcolor{blue}{the parameters $\hat{\alpha}_{im}^{(\tau_k)}$ and $\hat{\mathbf{B}}_{im}^{(\tau_k)}$} for each $\tau_k\in\mathbbm{Q}$; \textcolor{blue}{where $i$ and $m$ indicate that these parameters are associated with $y_{i,t}$ and using $m$ principle components, respectively.}
		
		Compute mean and covariance matrix for the quantiles as
		\begin{align*}
		\textcolor{blue}{\hat{\bm{\mu}}_i\leftarrow\left(\hat{\alpha}_{im}^{(\tau_k)}:\tau_k \in \mathbbm{Q}\right),\hat{\mathbf{V}}_{i}\leftarrow\left(\hat{\mathbf{B}}_{im}^{(\tau_{k})}\mbox{diag}(\lambda_{1},\ldots,\lambda_{m})\hat{\mathbf{B}}_{im}^{(\tau_{l})'}\right)_{kl}}
		\end{align*}
		
		\For{$b=1,\ldots,B$}{
			Draw one $q$-dimensional sample $\mathbf{q}_b\sim\mathcal{N}\left(\hat{\bm{\mu}}_i,\hat{\mathbf{V}}_i\right)$;
			
			Interpolate $\mathbf{q}_b$ through shape-preserving interpolation to a distribution $\hat{F}_{T+1}|\mathbf{q}_b$;
			
		}
		Sample from $\hat{F}_{T+1}|\mathbf{q}_1, \ldots, \hat{F}_{T+1}|\mathbf{q}_B$ and aggregate samples to an estimate of $\hat{F}_{T+1}$, the unconditional distribution function of $\mathbf{y}_{T+1}$ with an empirical distribution function;
	}
	Generate the multivariate distribution with the marginal distributions and a copula.
	\begin{align*}
	\hat{F}_{T+1}(\mathbf{y})\leftarrow C\left(\hat{F}_{T+1,1}(y_1),\ldots,\hat{F}_{T+1,n}(y_n) \right);
	\end{align*}
	\caption{Factor Quantile Model with Bootstrap Aggregation}
	\label{Algorithm:FactorQuantileAB}
\end{algorithm}

 We can also summarise the algorithm in the following stages:
\begin{description}
	\item[Stage 1] Given observations at times $t = 1, \ldots, T$ for $n$ stationary zero-mean stochastic variables $\mathbf{y}$ take the spectral decomposition of their covariance matrix $\bm{\Sigma}$ and thereby select the first $m$ principal components for common factors, denoted $\mathbf{x}_{m}$;
	\item[Stage 2] Using the same sample, estimate quantile regressions of the form \eqref{Equation:QuantileModel} for each $\tau$-quantile in turn, where $\tau\in(0,1)$ are pre-specified by a grid $\mathbbm{Q}$ of $(0,1)$;
	\item[Stage 3] We sample $\left(\hat{y}_{k}^{(\tau_1)},\ldots,\hat{y}_{k}^{(\tau_q)}\right)$ from the multivariate Gaussian distribution, whose mean and variance are specified in equations \eqref{Equation:Expectation} and \eqref{Equation:Covariance}, we then apply shape-preserving interpolation to construct a marginal distribution and then take a sample size $N$ from this distribution. Repeat this sampling $B$ times. These $N \times B$ observations are combined to form the conditional marginal distribution $\hat{f}_{k}$ for each element $k$ of the dependent variable;
	
	\item[Stage 4] Select a copula function to obtain the multivariate distribution forecast.
\end{description}

Bootstrap sampling is extremely fast so we can set $N$ and $B$ to be very large numbers. For instance, in the empirical study of the next section we set $N = 100{,}000$ and $B = 250$ so that 25 million samples are taken from each conditional marginal during the bagging algorithm. Alternative methods such as kernel density estimation could also be applied to aggregate the $N \times B$ observations to a distribution. However, this is not really necessary when $N \times B$ is so large.

A simple alternative to bagging is to use only the expected value \eqref{Equation:Expectation} and ignore the covariances \eqref{Equation:Covariance}. When the FQ models use the first few principal components, there exists  considerable variation about the point forecast, which is the unconditional expectation, of the quantiles, because the higher principal components have the greatest variation. So instead, this ``alpha'' latent FQ version employs the last $n-m$ principal components in quantile regressions. This way, the intercept captures a point forecast about which there is much less variation. The statistical properties described above remain valid as the common factors remain uncorrelated, but now the intercepts $\hat{\bm{\alpha}}^{(\tau)}_{n-m}$ capture an expected value with little variation: the covariance \eqref{Equation:Covariance} is minimal because $\left(\lambda_{m+1}, \ldots,\lambda_{n}\right)$ \textcolor{blue}{consists of the smallest eigenvalues}.  
This is not a new idea. Following \citet{jensen1968}, using the intercept to encompass the remaining variation not explained by factors is now widely applied to the performance evaluation of portfolio managers. 

\section{Empirical Results}
\label{Section:EmpiricalStudy}
   We compare FQ models with benchmark models that are commonly applied to systems of financial and economic variables: (i) two asymmetric Student-$t$ multivariate GARCH(1,1) models, and (ii) a Gaussian copula with empirical marginals. These have been selected as (i) the family of parametric dynamic models which best capture the salient properties of financial time-series i.e. volatility clustering, skew and heavy tails, asymmetric response to shocks, and (ii) a copula which is amenable to high-dimensional systems and also performs well in previous forecast exercises \citep{patton2012, patton2013}. Of course, there are a plethora of models available but including further models would provide so much information as to detract from the clear messages of this paper. Also note that, since GARCH models do not scale well to higher dimensions, we have limited the dimensions of the systems selected in our empirical study. That is the only reason we have not considered very large systems. FQ models scale easily and naturally to higher dimensions and retain very fast calibration times.

  Let $F$ denote the data generation process. A scoring rule is proper if the expected score is minimized when the forecaster issues the probabilistic forecast $F$, rather than another distribution $G \ne F$, and it is strictly proper if this minimum is unique -- see \citet{gneiting2007b} for further discussion. Since the goal of probabilistic forecast is to maximize sharpness of the distribution forecast, subject to calibration, we focus our assessment on proper scoring rules which address both calibration and sharpness simultaneously \citep{winkler1996}. 
Also, as recommended by \citet{gneiting2008b} and \citet{scheuerer2015} we utilize multiple univariate and multivariate proper scores. 

We compare two versions of our latent FQ model with two standard econometric models for predicting systems of exchange rates, term structures of interest rates and commodity future indices, assessing the model's accuracy using univariate and multivariate proper scoring rules. \autoref{Section:EmpiricalDesign} begins with a specification of the proper scoring rules and briefly describes the benchmark models. Then \autoref{Section:Data} details the data used for this empirical study and outlines the model calibration. \autoref{Section:UnivariateResults} presents results obtained using the weighted CRPS for univariate distribution forecasts and \autoref{Section:MultivariateResults} summarises results for multivariate distribution forecasts using the energy and variogram scores with different parameters. For reasons of space, many results cannot be reported in detail but they are available from the authors on request, along with the data and code used to generate these results. 

\subsection{Empirical Design}
\label{Section:EmpiricalDesign}
\textcolor{blue}{Scoring rules are a type of distance measure between
	a predictive distribution and an observation. As such they can be
	used to compare the predictive performance of competing models.} 
In the class of densities with finite first moments the weighted CRPS is a strictly proper scoring rule which is easy to compute and very flexible. It compares distribution forecasts by focussing on certain regions of interest, such as the centre or the tails. Introduced by \citet{matheson1976}, it is the recent work of \citet{gneiting2011} that really drew attention to this score, and the need for proper scoring rules applied to univariate distribution forecasting. Given a forecast distribution $F$ of an unknown data generation process and a realization $y$ from this unknown process, the weighted CRPS is defined as
\begin{align*}
    C_w(F,y) = 2 \int_{0}^{1} \left( \mathbbm{1}\{y \leq F^{-1}(\alpha)\}-\alpha \right) \left( F^{-1}(\alpha)-y \right) w(\alpha) \,\mbox{d}\alpha, 
\end{align*} 
where $w(\alpha)$ is a weight function which specifies a focus on particular parts of the distribution. \citet{gneiting2011} recommend using $w(\alpha)=1$ for the entire distribution, $w(\alpha)=\alpha(1-\alpha)$ for the centre, $w(\alpha)=\alpha^{2}$ for the left tail, $w(\alpha)=(1-\alpha)^{2}$ for the right tail and $w(\alpha)=(2\alpha-1)^{2}$ for both tails of the distribution. 

For ranking multivariate distribution forecasts with proper scoring rules we consider the energy score \citep{szekely2003} which generalizes the kernel representation of CRPS specified by \citet{gneiting2008b} and \textcolor{blue}{the variogram score is proposed by \citep{scheuerer2015} via a completely different construction principle}. To define these scores we require the following notation: Let $\mathbf{y}=(y_{1}, \ldots, y_{n})'$ be an observation of the $n$-variate random vector $\mathbf{Y}$ and let $F$ be a forecast of the multivariate distribution of $\mathbf{Y}$. The energy score is defined as 
\begin{align*}
	\textcolor{blue}{
    \mbox{ES}(F,\mathbf{y})=-\frac{1}{2}\mathbb{E}_F\left(\lVert\mathbf{Y}-\mathbf{Y}'\rVert\right)+\mathbb{E}_F\left(\lVert\mathbf{Y-\mathbf{y}}\rVert\right)}
\end{align*}
where $\lVert\cdot\rVert$ denotes the Euclidean norm and $\mathbf{Y}$ and $\mathbf{Y}'$ are independent random vectors with distribution $F\in\mathcal{F}$, the class of Borel probability measures such that $\mathbb{E}_F(\lVert\mathbf{Y}\rVert)$ is finite. \citet{szekely2003} proves that the energy score is strictly proper relative to $\mathcal{F}$. The variogram score of order $p$ is defined as
\begin{align*}
\mbox{VS}_p(F,\mathbf{y})=\sum^n_{i,j=1}\left(|y_i-y_j|^p-\mathbb{E}_F\left(|Y_i-Y_j|^p\right)\right)^2
\end{align*}
where $Y_i$ and $Y_j$ are the $i$-th and $j$-th component of a random vector with distribution $F$. The score is proper relative to the class of the probability distributions for which the $2p$-th moments of all components are finite. The inclusion of the variogram score is especially important since the energy score is not sensitive to misspecification of correlations 
\citep{pinson2012}.

To rank the performance of the competing models, we employ the Model Confidence Set (MCS) of \citet{hansen2011} based on the three proper scores above. Given a loss function and an initial set $\mathcal{M}^0$ containing all competing models, MCS applies a sequential equivalence test and an elimination rule to apply when this test is rejected. For some pre-specified $\alpha$, MCS returns a set of superior models $\mathcal{M}^*_{\alpha}$ that includes the best models in $\mathcal{M}^0$, in the sense that their performance cannot be distinguished with equivalence tests at a confidence level of $1-\alpha$.

Consider a finite set $\mathcal{M}$ with models indexed by $i=1,\ldots,N$ and a loss function $L$, so that $L_{it}$ is the loss of model $i$ for a forecast at time $t$. Then for $i,j=1,\ldots,N$ and $t=1,\ldots,T$ we define $d_{ij,t}:=L_{it}-L_{jt}$ and $\mu_{ij}:=\mathbb{E}(d_{ij,t})$. To test $\mbox{H}_0^{\mathcal{M}}:\mu_{ij}=0$ for all $i,j$ versus $\mbox{H}_A^{\mathcal{M}}:\mu_{ij}\neq0$ for some $i\ne j$ the MCS test statistic is $$T_{\mathcal{M}}:=\max_{i,j\in\mathcal{M}}\left|{\overline{d_{ij}}}/{\sqrt{\hat{\sigma}^2}}\right| \qquad \qquad \mbox{where} \qquad \qquad  \overline{d_{ij}}=T^{-1}\sum^T_{t=1}d_{ij,t},$$ and $\hat{\sigma}^2$ is the bootstrapped estimate of the variance of $d_{ij,t}$. Since the distributions of $T_{\mathcal{M}}$ are non-standard, they have to be estimated through a bootstrap procedure and, as suggested by \citet{hansen2011}, this should avoid the estimation of high-dimensional covariance matrices. To this end we employ a block–bootstrap where the block-length is determined by the maximum number of significant parameters during the fitting of an autoregressive model on the relative performance variable. 

If the hypothesis of equal predictive ability is rejected we then identify the worst model $e$ using the elimination rule $e=\argmax_{i}\left\{\sup_{j\in\mathcal{M}}{\overline{d_{ij}}}/{\sqrt{\hat{\sigma}^2}} \right\}$, and repeat the testing procedure with the updated model set $\mathcal{M}\backslash\{e\}$. Otherwise, we set $\mathcal{M}^*=\mathcal{M}$. This way, forecasting accuracy can be assessed by the frequency that each model remains in the final set $\mathcal{M}^*_{\alpha}$. The number of models in the MCS increases as we decrease $\alpha$, just like the size of a confidence interval. We follow \citet{hansen2011} and most empirical work since, using $\alpha = 0.25$ and $0.1$ to generate the 75\% and the 90\% MCS, the former being a sub-set of the latter.

Next we define the two classes of established models used in our study. The multivariate GARCH models are from the family of Constant Conditional Correlation GARCH ({CCC-GARCH}) of \citet{bollerslev1990} and the Dynamic Conditional Correlation GARCH ({DCC-GARCH}) family of \citet{engle2002}, both of which are widely used in literature. This choice is motivated by \citet{hansen2005b} who provide an extensive comparison of 330 univariate GARCH specifications, using the \citet{hansen2005a} superior predictive ability data-snooping check, concluding that it is hard to beat an asymmetric GARCH(1,1) model with Student-$t$ innovations. For some exchange rates the symmetric version is sufficient, and for some stocks a Gaussian conditional distribution for the errors performs as well, but these models are nested within our more general multivariate specification. Both CCC- and DCC-GARCH are based on the decomposition of the covariance matrix $\bm{\Sigma}_t$ of the asset returns with $\bm{\Sigma}_t=\mathbf{D}_t\mathbf{C}_t\mathbf{D}_t$, where $\mathbf{D}_t$ is a diagonal matrix of the time-varying univariate GARCH volatilities. To account for the well-documented asymmetric response of volatility to positive and negative shocks in returns we employ the E-GARCH model of \citet{nelson1991} with Student-$t$ innovations for the variances of each asset return $y_{it}$, thereby specifying the following data generation process:
\begin{equation}
    \begin{aligned}
        \label{Equation:EGARCH}
        y_{it}&= \mu+\varepsilon_{it},\quad \varepsilon_{it} =\sigma_{it} z^\nu_{it}, \\
        \log\left(\sigma_{it}^2\right)&=\kappa_i+\gamma_i \log\left(\sigma_{i,t-1}^2\right)+\alpha_i\left[\frac{|\varepsilon_{i,t-1}|}{\sigma_{i,t-1}}-\mathbb{E}\left(\frac{|\varepsilon_{i,t-1}|}{\sigma_{i,t-1}}\right)\right]+\xi_i\left(\frac{\varepsilon_{i,t-1}}{\sigma_{i,t-1}}\right), 
    \end{aligned}
\end{equation}
where $z^\nu_{it}$ follows a Student-$t$ distribution with $\nu_i$ degrees of freedom and $\kappa_i, \gamma_i, \alpha_i, \xi_i$ are the GARCH parameters. \citet{bollerslev1990} assumes that the correlation matrix $\mathbf{C}_t$ is not time varying and uses a constant correlation matrix in CCC-GARCH while \citet{engle2002} extends $\mathbf{C}_t$ in DCC-GARCH to a time-varying but non-stochastic matrix. The literature on empirical studies which compare the accuracy of different asymmetric univariate GARCH models does not agree on a single superior parametrisation. Therefore it is unlikely that our results would change if we employed a different asymmetric model (e.g. GJR-GARCH). Besides this, our purpose here is not to test the accuracy of different GARCH models, it is to validate the use of FQ models relative to the standard models that are commonly used for predicting financial returns. 

The other model class which has established itself in the finance literature are based on an Empirical Distribution Function (EDF). These will be combined into a multivariate distribution using a Gaussian copula with a historical correlation matrix estimated on the same data used for calibration. There are, of course, numerous alternative parametric choices for both marginals and copula, as described by \citet{patton2013}. \textcolor{blue}{We opted for the EDF because it is simple to implement in practice. In addition, it has been shown that the EDF approach performs competitively when compared with parametric approaches in the literature.} Using EDF marginals based on the same data as the FQ marginals allows us to test the effectiveness of PCA factor models, in the context of quantile regressions, for reducing the noisy variation which could deteriorate forecasting accuracy of models with EDF marginals. To summarize, we have selected a parsimonious set of alternative models and two different benchmark FQ parametrisations, and we shall also compare the performance of specific scoring rules and their ability to test our assumptions regarding both marginals and correlation structure.

\subsection{Data and Calibration}
\label{Section:Data}
We use three eight-dimensional multivariate time-series data sets -- on USD exchange rates, US interest rates and Bloomberg investable commodity indices. Within each set we have selected variables to broadly represent the asset class: the exchange rates are those with the highest trading volume (excluding the Chinese Renminbi, which was pegged to the USD until recently); the interest rates span the term structure of US Treasury bonds from 6 months to 20 years; and the commodity indices are chosen to represent the energy, metals, softs and livestock sectors. The exchange rate and commodity index data are obtained from Thomson Reuters Datastream and the interest rates data are downloaded from the US Treasury website. All series are daily and end on 31 December 2018 but the start date varies with data availability, being 01 January 1991 for the commodity indices, 01 January 1994 for the interest rates and 01 July 1999 for the exchange rates. The exchange rate data started in 1999 and we need 2000 observations to calibrate the models. With approximately 250 trading days per year, we could not start the exchange rate evaluations until March 2007. We used the same sub-samples for all data sets, for comparison. The models are re-calibrated daily and the estimated parameters are used to generate one-day-ahead distribution forecasts. Then the fixed-size calibration sample is rolled forward one day and the forecasts are repeated. We avoid data snooping by using a broad range of data sets with assets motivated through economic factors rather than the predictive prowess of our models. All parameters of the FQ models are chosen based on criteria that are available ex-ante. Additionally, we quantify the performance based on very long time-series, further limiting the probability that any superior performance can be attributed to chance.

We calibrate the multivariate GARCH models using maximum likelihood estimators adapted from the implementation in the Oxford MFE Toolbox by \citet{MFEToolbox2013} to utilize E-GARCH with Student-$t$ distributed innovations. That is, we have replaced the univariate Gaussian GARCH(1,1) in the MFE toolbox code for CCC- and DCC-GARCH with Student-$t$ E-GARCH. It is well known that multivariate GARCH models can have ill-conditioned likelihood functions which are hard to optimize unless the calibration sample has sufficient size, so we have selected daily 2,000 returns for each time-series for this calibration. We prefer to confine each set to eight dimensions to limit the computational complexity when estimating the multivariate GARCH models. This point is discussed in more detail at the end of this sub-section. 

Regarding the FQ specifications, we apply the latent versions based on the last principal components (FQ-A) and bagging (FQ-B) with the same Gaussian copula as the EDF. Both specifications of our FQ model use the quantile  grid \textcolor{blue}{of quantile levels} $\mathbbm{Q}_9$ from \autoref{Section:Example} for the regressions and employ the shape-preserving method for interpolating distribution functions.    The estimated conditional quantiles exhibited no crossing behaviour on any data set with any of the calibration choices in our empirical study, indicating that our factor models are well-conditioned. However, if any issues with non-monotonicity of quantiles do arise in an application a number of methods could be employed to ensure non-crossing quantiles. For instance, see \citet{koenker2005}, \citet{chernozhukov2010}, \citet{Rodrigues2017} or \citet{Santos2020}.

\begin{figure}[!htb]
    \caption{Cumulative variance explained by the principal components}
     \vspace{-0.3cm}
        \caption*{\footnotesize   The variance explained by each principal component is derived by applying PCA on daily rolling windows of 250 data points. Each data set starts on a different date and the results begin approximately one year after the start date.}
    \label{Figure:VarianceExplained}
    \centering
    \includegraphics[width=\textwidth]{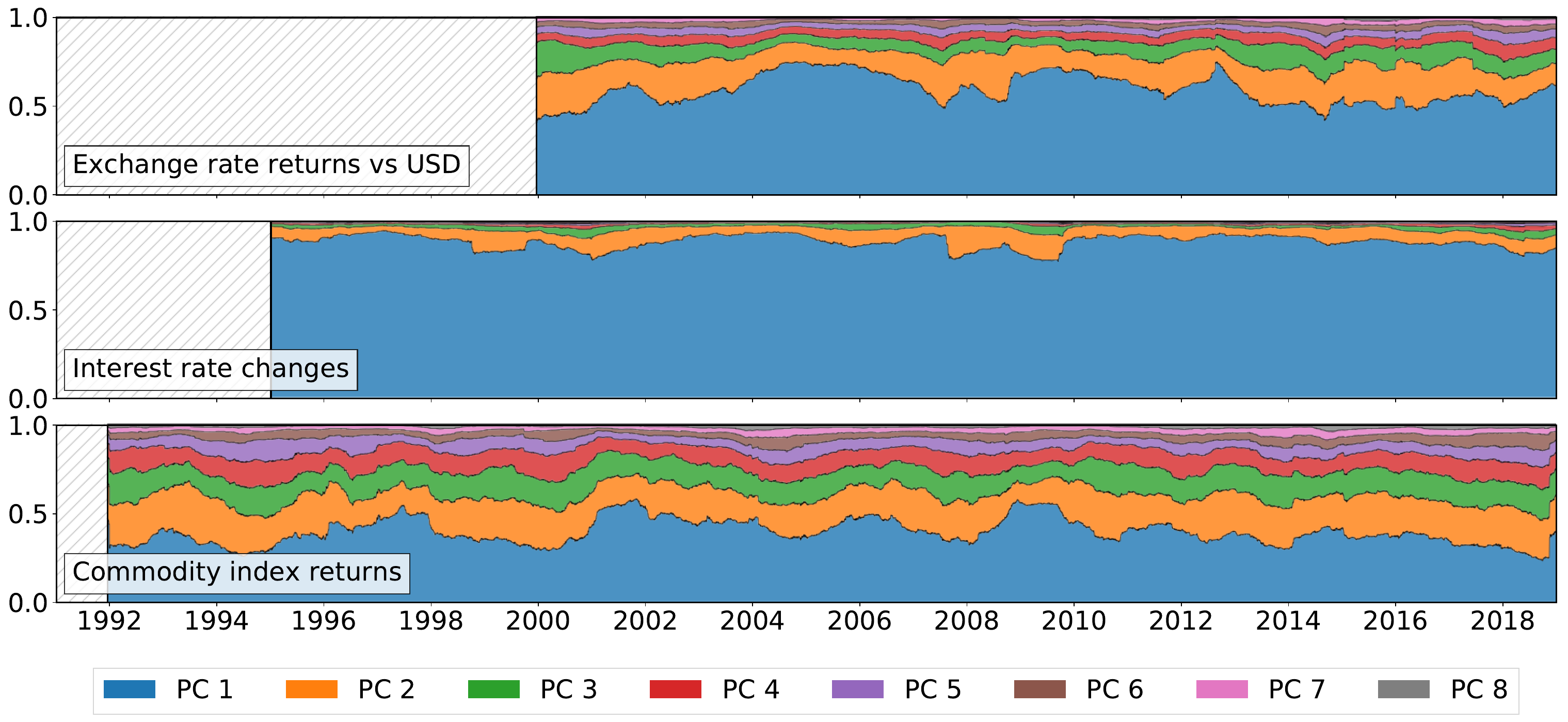}
\end{figure}

In FQ-B, we select $m=4$ components as common factors for the exchange rates, $m=2$ for the interest rates and $m=6$ for the commodity indices. By depicting the cumulative variance explained by the rolling principal components over the available data period for each asset class, \autoref{Figure:VarianceExplained} motivates how these values of $m$ are selected. On average, over the entire period shown, together the four components explain 90\% of the variation in the exchange rate data, the two components explain 95\% of the variation in the interest rates, and the six components explain 95\% of the variation in the commodity returns. Following the same reasoning, FQ-A uses $m=4$ components as common factors for the exchange rates, $m=6$ for the interest rates and $m=2$ for the commodity indices. 

Both GARCH models are restricted to long calibration periods for the estimation of the long-term variance and the stability of calibrated parameters. For consistent comparison with the GARCH models, which are not well-conditioned on smaller \textcolor{blue}{calibration sizes}, we have also taken 2,000 data points for the other models. But EDF and FQ models are likely to perform better on smaller \textcolor{blue}{calibration sizes}. With principal component factors FQ models yield robust estimates even with a \textcolor{blue}{calibration size} of 250. Here we present results for the FQ and EDF models only for \textcolor{blue}{calibration sizes} of 2,000 and 250 although others are available on request.\footnote{We compared several \textcolor{blue}{calibration sizes} between 250 and 2000 finding that, in general, \textcolor{blue}{the models with smaller calibration sizes performed better than those with larger ones}. We included a \textcolor{blue}{calibration} sample of size 2000 for comparison with the GARCH models, because these could not be calibrated robustly on a \textcolor{blue}{calibration size} of 250. }\autoref{Table:ModelDescription} summarises the models that we apply in the remainder of this study.

\begin{table}[!htb]
	\centering
	\small
	\caption{Summary of models used in the empirical study}
	     \vspace{-0.3cm}
	\caption*{\footnotesize {   This table summarizes the acronyms used to denote each model: FQ stands for factor quantile; EDF stands for empirical distribution function; CCC stands for  constant conditional correlation -- which is equivalent to the Gaussian copula calibrated to a sample size 2000; and DCC stands for the time-varying dynamic conditional correlation matrix}}
	\label{Table:ModelDescription}
	\begin{tabular}[!htb]{lllr}
		\midrule
		\midrule
		Model & Marginals & Dependency & Calibration size\\
		\midrule
			{FQ-A}$_{250}$  & Alpha FQ            & Gaussian copula                & 250   \\
		{FQ-A}$_{2000}$ & Alpha FQ          & Gaussian copula                & 2,000 \\
		{FQ-B}$_{250}$  & Bagging FQ              & Gaussian copula                & 250   \\
		{FQ-B}$_{2000}$ &  Bagging FQ              & Gaussian copula                & 2,000 \\
		{EDF}$_{250}$    & EDF                           & Gaussian copula                & 250   \\
		{EDF}$_{2000}$   & EDF                           & Gaussian copula                & 2,000 \\
		CCC-GARCH     & Student-t E-GARCH(1,1)        & CCC        & 2,000 \\
		DCC-GARCH     & Student-t E-GARCH(1,1)        & DCC        & 2,000 \\
		\midrule
	\end{tabular}
\end{table}
 
\begin{figure}[!htb]
    \caption{Convergence issues with GARCH models (sugar)}
     \vspace{-0.3cm}
        \caption*{\footnotesize The parameter illustrated is the constant estimated for the sugar marginal in DCC-GARCH. The upper figure shows the parameter obtained using the adapted Oxford MFE toolbox and the lower figure shows the parameter after replacing erroneous calibrations with the most recent unproblematic value. Parameters that differ by a very large amount from previous estimations are classified as mis-calibrations   and are marked by red crosses in the upper figure.} 
    \label{Figure:DCCConvergence}
    \centering
    \includegraphics[width=\textwidth]{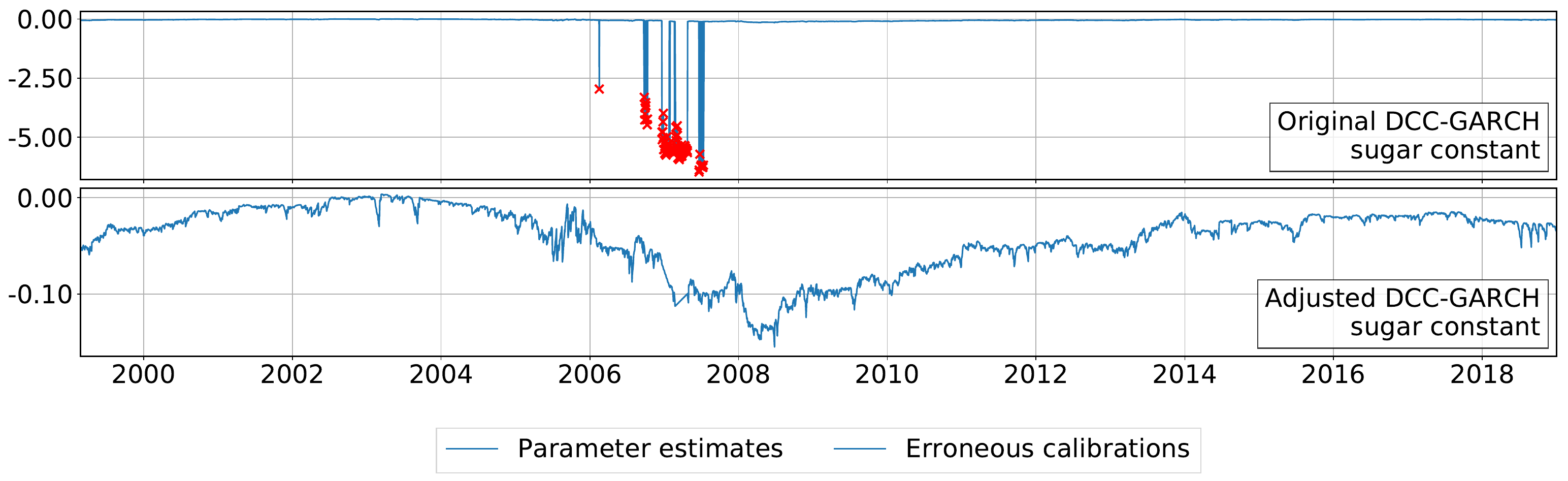}
\end{figure}

Despite the necessarily large \textcolor{blue}{calibration sizes}, both multivariate GARCH models exhibit calibration issues because, for an eight dimensional time-series, there are (at least) 40 parameters so the likelihood surfaces are very challenging to optimize.  Sometimes parameter estimates do not converge to sensible values, particularly for the commodities data, and in such cases we exchange erroneous parameters with the most recent unproblematic values, as illustrated by \autoref{Figure:DCCConvergence}. To locate and correct mis-calibrated parameters requires manual attention, which prevents full automation of multivariate GARCH models. 
 
We emphasise that FQ models are much easier to fully automate and by no means limited to eight dimensions. Thus, they are more amendable to the type of high-frequency trading that is common among hedge funds that often employ algorithms to re-balance portfolios every day. The only reason that we have restricted this study to eight dimensions is that the problems documented above with calibrating GARCH parameters are even further exacerbated. Further note that FQ models are much faster to calibrate than multivariate GARCH, even without dealing with any of the latter's convergence issues. In our empirical study the FQ models were over 30\% faster than CCC-GARCH and more than five times faster than DCC-GARCH. Also, our implementation of FQ models is based on Python while the multivariate GARCH models use optimized MATLAB functions. As the efficiency of MATLAB is generally higher than that of Python scripts, we expect that the difference in speed would become even more pronounced when comparing the multivariate GARCH models to an optimized FQ algorithm.

\subsection{Univariate Forecasting Accuracy Results}
\label{Section:UnivariateResults}
Both \citet{gneiting2008b} and \citet{scheuerer2015} emphasise the importance of testing the accuracy of univariate distribution forecasts derived from multivariate models. Applying multivariate tests alone is not sufficient because we require a model that forecasts accurate marginals as well as one that correctly captures the dependence between them. So in this section we present the results of applying weighted CRPS to each model listed in the initial set $\mathcal{M}_0$ defined in \autoref{Table:ModelDescription} and then finding the MCS derived from these scores. 

\begin{table}[!htb]
	\small
	\centering{}\textcolor{blue}{\setlength{\tabcolsep}{1.2pt} \centering
		\caption{Average values of uniformly-weighted CRPS.} 
		  \vspace{-0.3cm}
		\caption*{ The  table reports average uniformly weighted CRPS over the entire out-of-sample period for each model and each univariate data set. The \textcolor{blue}{models} having the lowest average \textcolor{blue}{scores} are highlighted in blue. Beside the score we indicate which models are in the 90\% and 75\% MCS, using {*}
		and {*}{*}, respectively, the latter being a sub-set of the former. Note that when there is only one superior model in the 75\% MCS, this will also be in the 90\%  MCS. 
		The bottom row summarizes the percentages each model is in the 90\%
			MCS for all 24 data sets. That is, the number of {*} or {**} occurrences in each column, divided by 24.}
		\vspace{-0.1cm}
		\label{Table: raw CRPS} }%
	\begin{tabular}[t]{L{2cm}llllllll}
		\toprule 
		\multirow{1}{*}{Asset} & \multicolumn{2}{c}{$\mbox{GARCH}$} & \multicolumn{2}{c}{$\mbox{EDF}$} & \multicolumn{2}{c}{$\mbox{FQ-B}$} & \multicolumn{2}{l}{$\mbox{FQ-A}$}\tabularnewline
		& \multicolumn{1}{l}{CCC} & \multicolumn{1}{l}{DCC} & \multicolumn{1}{l}{$250$} & \multicolumn{1}{l}{$2000$} & \multicolumn{1}{l}{$250$} & $2000$ & $250$ & $2000$\tabularnewline
		\midrule 
		\multicolumn{4}{l}{\textit{Exchange rate returns ($\times10^{-3}$)}} &  &  &  &  & \tabularnewline
		AUD & 4.46 & \hl{$4.37^{**}$} & 4.46 & 4.49 & 4.45 & 4.55 & 4.43 & 4.49\tabularnewline
		CAD & \hl{$3.25^{**}$} & 3.26 & 3.32 & 3.36 & 3.30 & 3.40 & 3.31 & 3.36\tabularnewline
		CHF & 3.63 & \hl{$3.48^{**}$} & 3.52 & 3.53 & 3.56 & 3.57 & 3.50 & 3.52\tabularnewline
		EUR & $3.35^{**}$ & 3.58 & 3.36 & 3.37 & $3.35^{**}$ & 3.40 & \hl{$3.34^{**}$} & $3.36^{**}$\tabularnewline
		GBP & 3.89 & \hl{$3.18^{**}$} & 3.24 & 3.25 & 3.24 & 3.28 & 3.25 & 3.29\tabularnewline
		JPY & $3.50^{**}$ & 3.53 & 3.52 & 3.53 & 3.52 & 3.57 & \hl{$3.49^{**}$} & 3.53\tabularnewline
		NZD & \hl{$4.55^{**}$} & 4.74 & 4.66 & 4.70 & 4.64 & 4.73 & 4.64 & 4.70\tabularnewline
		SEK & \hl{$4.27^{**}$} & 4.79 & 4.31 & 4.35 & $4.30^{*}$ & 4.39 & $4.28^{**}$ & 4.35\tabularnewline
		&  &  &  &  &  &  &  & \tabularnewline
		\multicolumn{3}{l}{\textit{Interest rate changes}} &  &  &  &  &  & \tabularnewline
		6 month & \hl{$1.33^{**}$} & 1.37 & 1.37 & 1.44 & 1.37 & 1.47 & 1.37 & 1.47\tabularnewline
		1 year & 3.09 & 2.91 & \hl{$1.55^{**}$} & 1.63 & 1.56 & 1.63 & 1.56 & 1.64\tabularnewline
		2 year & 2.68 & 2.67 & \hl{$2.42$} & 2.50 & \hl{$2.42^{**}$} & 2.50 & \hl{$2.42^{*}$} & 2.51\tabularnewline
		3 year & 3.12 & 3.23 & 2.76 & 2.82 & \hl{$2.75^{**}$} & 2.82 & \hl{$2.75^{**}$} & 2.83\tabularnewline
		5 year & 3.17 & 3.19 & 3.16 & 3.20 & \hl{$3.14^{**}$} & 3.20 & \hl{$3.14^{**}$} & 3.21\tabularnewline
		7 year & 3.49 & 3.51 & 3.26 & 3.29 & \hl{$3.24^{**}$} & 3.29 & \hl{$3.24^{**}$} & 3.30\tabularnewline
		10 year & 3.51 & 3.51 & 3.11 & 3.14 & \hl{$3.10^{**}$} & 3.14 & \hl{$3.10^{**}$} & 3.14\tabularnewline
		20 year & 3.47 & 3.51 & 3.01 & 3.03 & \hl{$2.99^{**}$} & 3.03 & \hl{$2.99^{**}$} & 3.03\tabularnewline
		&  &  &  &  &  &  &  & \tabularnewline
		\multicolumn{4}{l}{\textit{Commodity index returns ($\times10^{-3}$)}} &  &  &  &  & \tabularnewline
		Copper & $8.60^{*}$ & \hl{$8.59^{**}$} & 8.68 & 8.81 & 8.67 & 8.92 & $8.62^{*}$ & 8.81\tabularnewline
		Corn & \hl{$8.57^{**}$} & 8.95 & 8.81 & 8.88 & 8.82 & 8.95 & 8.77 & 8.89\tabularnewline
		Gold & 6.21 & 5.84 & \hl{$5.70^{**}$} & 5.74 & 5.74 & 5.80 & 5.83 & 5.90\tabularnewline
		Live Cattle & 5.06 & \hl{$4.99^{**}$} & 5.05 & 5.07 & 5.05 & 5.10 & 5.45 & 5.56\tabularnewline
		Nat. Gas & 15.16 & \hl{$14.98^{**}$} & 15.25 & 15.26 & 15.22 & 15.37 & 15.15 & 15.25\tabularnewline
		Soybean & 8.04 & \hl{$7.68^{**}$} & 7.85 & 7.88 & 7.84 & 7.95 & 7.83 & 7.88\tabularnewline
		Sugar & 10.98 & 10.96 & 10.93 & 11.02 & 10.94 & 11.08 & \hl{$10.86^{**}$} & 11.01\tabularnewline
		WTI Oil & \hl{$11.37^{**}$} & 11.66 & 11.46 & 11.53 & 11.44 & 11.60 & $11.38^{**}$ & 11.52\tabularnewline
		&  &  &  &  &  &  &  & \tabularnewline
		\textit{Summary} & 37.5\% & 29.2\% & 8.3\% & 0.0\% & 33.3\% & 0.0\% & 50.0\% & 4.2\%\tabularnewline
		\bottomrule
	\end{tabular}\textcolor{blue}{\small{}\hspace{0.1cm}}
\end{table}

\begin{table}[!htb]
	\small
	\centering{}\textcolor{blue}{\setlength{\tabcolsep}{1.2pt} \centering
		\caption{Average values of centre-weighted CRPS.}  
			  \vspace{-0.3cm}
			\caption*{\footnotesize  The  table reports average centre weighted CRPS over the entire out-of-sample period for each model and each univariate data set. The \textcolor{blue}{models} having the lowest average \textcolor{blue}{scores} are highlighted in blue. Beside the score we indicate which models are in the 90\% and 75\% MCS, using {*}and {*}{*}, respectively, the latter being a sub-set of the former. Note that when there is only one superior model in the 75\% MCS, this will also be in the 90\%  MCS. 	The bottom row summarizes the percentages each model is in the 90\%	MCS for all 24 data sets. That is, the number of {*} or {**} occurrences in each column, divided by 24.}
			\vspace{-0.1cm}
		\label{Table: raw CRPS-1}} %
	\begin{tabular}[t]{L{2cm}llllllll}
		\toprule 
		\multirow{1}{*}{Asset} & \multicolumn{2}{c}{$\mbox{GARCH}$} & \multicolumn{2}{c}{$\mbox{EDF}$} & \multicolumn{2}{c}{$\mbox{FQ-B}$} & \multicolumn{2}{l}{$\mbox{FQ-A}$}\tabularnewline
		& \multicolumn{1}{l}{CCC} & \multicolumn{1}{l}{DCC} & \multicolumn{1}{l}{$250$} & \multicolumn{1}{l}{$2000$} & \multicolumn{1}{l}{$250$} & $2000$ & $250$ & $2000$\tabularnewline
		\midrule 
		\multicolumn{4}{l}{\textit{Exchange rate returns ($\times10^{-5}$)}} &  &  &  &  & \tabularnewline
		AUD & 85.83 & \hl{$84.97^{**}$} & 85.94 & 86.20 & 85.84 & 87.09 & $85.44^{*}$ & 86.13\tabularnewline
		CAD & \hl{$63.38^{**}$} & 63.50 & 64.31 & 64.70 & 64.04 & 65.39 & 64.05 & 64.65\tabularnewline
		CHF & $68.93$ & \hl{$67.43^{**}$} & 67.96 & 68.02 & 68.70 & 68.80 & $67.64^{*}$ & 67.93\tabularnewline
		EUR & 64.85 & 67.26 & 65.12 & 65.16 & $64.87^{**}$ & 65.70 & \hl{$64.80^{**}$} & $65.04^{**}$\tabularnewline
		GBP & 72.05 & \hl{$61.82^{**}$} & 62.58 & 62.71 & 62.63 & 63.18 & 62.58 & 63.12\tabularnewline
		JPY & $67.58^{**}$ & $68.01$ & 67.79 & 67.90 & 67.79 & 68.63 & \hl{$67.36^{**}$} & 67.85\tabularnewline
		NZD & \hl{$88.96^{**}$} & 91.30 & 90.50 & 90.78 & 90.04 & 91.29 & 90.08 & 90.71\tabularnewline
		SEK & \hl{$82.93^{**}$} & 90.28 & 83.48 & 83.90 & $83.37$ & 84.57 & $83.06^{**}$ & 83.89\tabularnewline
		&  &  &  &  &  &  &  & \tabularnewline
		\multicolumn{3}{l}{\textit{Interest rate changes ($\times10^{-2}$)}} &  &  &  &  &  & \tabularnewline
		6 month & \hl{$25.69^{**}$} & 26.07 & $25.91^{*}$ & 26.76 & 26.05 & 27.31 & 26.06 & 27.36\tabularnewline
		1 year & 54.12 & 51.17 & \hl{$29.76^{**}$} & 30.69 & 29.85 & 30.90 & 29.86 & 30.98\tabularnewline
		2 year & 49.72 & 49.55 & 46.88 & 47.86 & \hl{$46.81^{**}$} & 48.05 & 46.86 & 48.11\tabularnewline
		3 year & 58.83 & 60.54 & 53.59 & 54.41 & \hl{$53.40^{**}$} & 54.41 & $53.42^{**}$ & 54.49\tabularnewline
		5 year & 61.51 & 61.74 & 61.42 & 61.95 & \hl{$61.11^{**}$} & 61.97 & $61.12^{**}$ & 61.99\tabularnewline
		7 year & 65.96 & 66.18 & 63.42 & 63.81 & $63.11^{**}$ & 63.85 & \hl{$63.09^{**}$} & 63.85\tabularnewline
		10 year & 64.85 & 64.86 & 60.61 & 60.90 & $60.33^{**}$ & 60.99 & \hl{$60.33^{**}$} & 60.96\tabularnewline
		20 year & 63.38 & 63.74 & 58.55 & 58.77 & \hl{$58.28^{**}$} & 58.87 & $58.29^{**}$ & 58.84\tabularnewline
		&  &  &  &  &  &  &  & \tabularnewline
		\multicolumn{4}{l}{\textit{Commodity index returns ($\times10^{-4}$)}} &  &  &  &  & \tabularnewline
		Copper & $16.64^{**}$ & \hl{$16.63^{**}$} & 16.73 & 16.88 & 16.73 & 17.11 & $16.65^{**}$ & 16.88\tabularnewline
		Corn & \hl{$16.72^{**}$} & 17.22 & 17.03 & 17.10 & 17.05 & 17.24 & 16.96 & 17.10\tabularnewline
		Gold & 11.71 & 11.15 & \hl{$10.95^{**}$} & 11.01 & 11.03 & 11.13 & 11.06 & 11.15\tabularnewline
		Live Cattle & 9.80 & \hl{$9.71^{**}$} & 9.78 & 9.80 & 9.79 & 9.84 & 10.16 & 10.28\tabularnewline
		Nat. Gas & 29.41 & \hl{$29.18^{**}$} & 29.53 & 29.52 & 29.49 & 29.73 & 29.37 & 29.51\tabularnewline
		Soybean & 15.44 & \hl{$14.95^{**}$} & 15.17 & 15.18 & 15.16 & 15.32 & 15.12 & 15.18\tabularnewline
		Sugar & 21.22 & 21.20 & 21.19 & 21.28 & 21.22 & 21.40 & \hl{$21.07^{**}$} & 21.28\tabularnewline
		WTI Oil & $22.06^{**}$ & 22.46 & 22.18 & 22.25 & 22.16 & 22.38 & \hl{$22.05^{**}$} & 22.24\tabularnewline
		&  &  &  &  &  &  &  & \tabularnewline
		\textit{Summary} & 37.5\% & 29.2\% & 12.5\% & 0.0\% & 29.2\% & 0.0\% & 54.2\% & 4.2\%\tabularnewline
		\bottomrule
	\end{tabular}\textcolor{blue}{\small{}\hspace{0.1cm}}
\end{table}

\begin{table}[!htb]
	\small
	\centering{}\textcolor{blue}{\setlength{\tabcolsep}{1.2pt} \centering
		\caption{ Average values of both-tail-weighted CRPS.} 
			  \vspace{-0.3cm}
			\caption*{\footnotesize The  table reports average both-tail weighted CRPS over the entire out-of-sample period for each model and each univariate data set. The \textcolor{blue}{models} having the lowest average scores are highlighted in blue. Beside the score we indicate which models are in the 90\% and 75\% MCS, using {*}
				and {*}{*}, respectively, the latter being a sub-set of the former. Note that when there is only one superior model in the 75\% MCS, this will also be in the 90\%  MCS. 
				The bottom row summarizes the percentages each model is in the 90\%
				MCS for all 24 data sets. That is, the number of {*} or {**} occurrences in each column, divided by 24.}
			\vspace{-0.1cm}
		\label{Table: raw CRPS-2} }%
	\begin{tabular}[t]{L{2cm}llllllll}
	\toprule 
	\multirow{1}{*}{Asset} & \multicolumn{2}{c}{$\mbox{GARCH}$} & \multicolumn{2}{c}{$\mbox{EDF}$} & \multicolumn{2}{c}{$\mbox{FQ-B}$} & \multicolumn{2}{l}{$\mbox{FQ-A}$}\tabularnewline
	& \multicolumn{1}{l}{CCC} & \multicolumn{1}{l}{DCC} & \multicolumn{1}{l}{$250$} & \multicolumn{1}{l}{$2000$} & \multicolumn{1}{l}{$250$} & $2000$ & $250$ & $2000$\tabularnewline
	\midrule 
	\multicolumn{4}{l}{\textit{Exchange rate returns ($\times10^{-3}$)}} &  &  &  &  & \tabularnewline
	AUD & 10.24 & \hl{$9.74^{**}$} & 10.21 & 10.45 & 10.14 & 10.63 & 10.12 & 10.45\tabularnewline
	CAD & \hl{$7.13^{**}$} & 7.19 & 7.47 & 7.75 & 7.42 & 7.86 & 7.46 & 7.69\tabularnewline
	CHF & 8.69 & \hl{$7.83^{**}$} & 8.02 & 8.07 & 8.15 & 8.21 & 7.96 & 8.05\tabularnewline
	EUR & 7.60 & 8.94 & 7.56 & 7.65 & $7.51^{**}$ & 7.74 & \hl{$7.50^{**}$} & 7.62\tabularnewline
	GBP & 10.05 & \hl{$7.10^{**}$} & 7.33 & 7.46 & 7.34 & 7.55 & 7.42 & 7.64\tabularnewline
	JPY & $8.00^{**}$ & 8.12 & 8.06 & 8.14 & 8.05 & 8.27 & $7.97^{**}$ & 8.13\tabularnewline
	NZD & $9.95^{**}$ & 10.92 & 10.45 & 10.68 & 10.34 & 10.79 & 10.36 & 10.68\tabularnewline
	SEK & $9.49^{**}$ & 11.80 & 9.67 & 9.97 & $9.61^{**}$ & 10.09 & $9.59^{**}$ & 9.97\tabularnewline
	&  &  &  &  &  &  &  & \tabularnewline
	\multicolumn{3}{l}{\textit{Interest rate changes}} &  &  &  &  &  & \tabularnewline
	6 month & \hl{$30.49^{**}$} & 32.34 & 32.94 & 36.92 & 32.95 & 37.48 & 32.91 & 37.79\tabularnewline
	1 year & 92.42 & 86.57 & $36.25^{**}$ & 39.90 & $36.21^{**}$ & 39.76 & \hl{$36.19^{**}$} & 40.20\tabularnewline
	2 year & 69.32 & 68.39 & 54.78 & 58.31 & $54.55^{**}$ & 58.17 & \hl{$54.55^{**}$} & 58.58\tabularnewline
	3 year & 76.60 & 81.08 & 61.86 & 64.84 & $61.48^{**}$ & 64.54 & \hl{$61.41^{**}$} & 64.93\tabularnewline
	5 year & 70.80 & 71.72 & 70.36 & 72.50 & $69.78^{**}$ & 72.38 & \hl{$69.73^{**}$} & 72.57\tabularnewline
	7 year & 85.32 & 86.68 & 72.25 & 74.07 & $71.64^{**}$ & 74.01 & \hl{$71.63^{**}$} & 74.10\tabularnewline
	10 year & 91.83 & 91.96 & 68.84 & 70.42 & $68.34^{**}$ & 70.41 & \hl{$68.26^{**}$} & 70.47\tabularnewline
	20 year & 93.66 & 96.00 & 66.41 & 67.74 & $65.95^{**}$ & 67.78 & \hl{$65.89^{**}$} & 67.78\tabularnewline
	&  &  &  &  &  &  &  & \tabularnewline
	\multicolumn{4}{l}{\textit{Commodity index returns ($\times10^{-3}$)}} &  &  &  &  & \tabularnewline
	Copper & 19.47 & $19.36^{**}$ & 19.83 & 20.54 & 19.78 & 20.81 & 19.65 & 20.54\tabularnewline
	Corn & $18.82^{**}$ & 20.61 & 20.00 & 20.39 & 19.98 & 20.59 & 19.86 & 20.49\tabularnewline
	Gold & 15.27 & 13.76 & \hl{$13.20^{**}$} & 13.40 & $13.27^{*}$ & 13.52 & 14.05 & 14.43\tabularnewline
	Live Cattle & 11.38 & \hl{$11.09^{**}$} & 11.37 & 11.50 & 11.37 & 11.59 & 13.90 & 14.49\tabularnewline
	Nat. Gas & 33.97 & \hl{$33.10^{**}$} & 34.35 & 34.46 & 34.27 & 34.74 & 34.03 & 34.44\tabularnewline
	Soybean & 18.64 & \hl{$16.97^{**}$} & 17.85 & 18.05 & 17.78 & 18.22 & 17.80 & 18.03\tabularnewline
	Sugar & 24.91 & 24.82 & 24.53 & 25.04 & 24.48 & 25.13 & \hl{$24.28^{**}$} & 25.00\tabularnewline
	WTI Oil & \hl{$25.48^{**}$} & 26.78 & 25.87 & 26.27 & 25.73 & 26.42 & $25.60^{**}$ & 26.25\tabularnewline
	&  &  &  &  &  &  &  & \tabularnewline
	\textit{Summary} & 29.2\% & 29.2\% & 8.3\% & 0.0\% & 41.7\% & 0.0\% & 50.0\% & 0.0\%\tabularnewline
	\bottomrule
\end{tabular}\textcolor{blue}{\small{}\hspace{0.1cm}}
\end{table}
  {\autoref{Table: raw CRPS}, \autoref{Table: raw CRPS-1} and \autoref{Table: raw CRPS-2} present the average scores for the entire out-of-sample period when the CRPS is uniformly weighted, centre weighted and both-tail weighted, respectively.  Smaller average scores are preferred. In each table  the model with lowest average score has this score highlighted in blue. In some cases the average scores differ only in the third decimal place, but this can still make a difference to inclusion in the MCS --- see, for instance, the results for 2-year interest rates in  \autoref{Table: raw CRPS}. We use one asterisk to indicate which models  are included in the 90\% MCS and two asterisks for models in the 75\% MCS. The latter could include more than one model but most tests identify a single model as the superior one. This suggests that our out-of-sample period is informative enough to select a best model unequivocally.  	For each of the 24 univariate data sets, the model(s) having lowest average score are highlighted in blue and the last row of each table reports the number of times a given model is included in the 90\% MCS, divided by 24.}

\begin{table}[!htb]
	\small
	\setlength{\tabcolsep}{3pt} \centering \caption{Comparison of univariate performance over time}
	\vspace{-0.3cm}
	\caption*{\footnotesize {This table shows the proportion of cases that each
			model is included in the 90\% MCS for each data set, over different sample periods and based on different weighting for CRPS. In each case this proportion is derived by counting the number of times the model is in the 90\% MCS and dividing this by 8, since there are 8 variables in each asset class. The model that is included in most of the MCS, for each asset class and sample period, and for each CRPS weighting, is  highlighted in
			blue.}}
	\label{Table:univariateMCSALSub} %
	\begin{tabular}[t]{llllllllll}
		\toprule 
		\multirow{1}{*}{} &  & \multicolumn{2}{c}{\hspace{-0.05cm}$\mbox{GARCH}$} & \multicolumn{2}{c}{\hspace{-0.05cm}$\mbox{EDF}$} & \multicolumn{2}{c}{\hspace{0cm}$\mbox{FQ-B}$} & \multicolumn{2}{l}{$\mbox{FQ-A}$}\tabularnewline
		&  & \multicolumn{1}{l}{CCC} & \multicolumn{1}{c}{DCC} & \multicolumn{1}{l}{$250$} & \multirow{1}{*}{$2000$} & \multicolumn{1}{c}{$250$} & $2000$ & $250$ & $2000$\tabularnewline
		\midrule 
		\multicolumn{2}{l}{\textit{Uniform}} &  &  &  &  &  &  &  & \tabularnewline
		& \textit{Exchange rate returns} &  &  &  &  &  &  &  & \tabularnewline
		& 2007 to 2010 & \hl{0.625} & 0.375 & 0.25 & 0.25 & 0.25 & 0.25 & 0.25 & 0.25\tabularnewline
		& 2011 to 2014 & 0.125 & 0.00 & 0.125 & 0.00 & 0.75 & 0.00 & \hl{0.875} & 0.00\tabularnewline
		& 2015 to 2018 & 0.50 & \hl{0.625} & 0.25 & 0.25 & 0.25 & 0.125 & 0.25 & 0.00\tabularnewline
		& \textit{Interest rate changes} &  &  &  &  &  &  &  & \tabularnewline
		& 2007 to 2010 & \hl{0.75} & 0.50 & 0.00 & 0.00 & 0.25 & 0.00 & 0.50 & 0.00\tabularnewline
		& 2011 to 2014 & 0.125 & 0.125 & \hl{0.50} & 0.00 & \hl{0.50} & 0.00 & \hl{0.50} & 0.00\tabularnewline
		& 2015 to 2018 & 0.375 & \hl{0.75} & 0.25 & 0.25 & 0.25 & 0.00 & 0.375 & 0.125\tabularnewline
		& \textit{Commodity index returns} &  &  &  &  &  &  &  & \tabularnewline
		& 2007 to 2010 & 0.50 & 0.375 & 0.00 & 0.25 & 0.625 & 0.00 & \hl{0.75} & 0.50\tabularnewline
		& 2011 to 2014 & 0.125 & 0.125 & 0.375 & 0.125 & \hl{0.625} & 0.25 & \hl{0.625} & 0.00\tabularnewline
		& 2015 to 2018 & 0.50 & \hl{0.75} & 0.125 & 0.125 & 0.50 & 0.125 & 0.375 & 0.25\tabularnewline
		&  &  &  &  &  &  &  &  & \tabularnewline
		\multicolumn{2}{l}{\textit{Centre }} &  &  &  &  &  &  &  & \tabularnewline
		& \textit{Exchange rate returns} &  &  &  &  &  &  &  & \tabularnewline
		& 2007 to 2010 & \hl{0.625} & 0.375 & 0.25 & 0.25 & 0.25 & 0.25 & 0.25 & 0.25\tabularnewline
		& 2011 to 2014 & 0.125 & 0.00 & 0.125 & 0.125 & \hl{0.875} & 0.00 & \hl{0.875} & 0.00\tabularnewline
		& 2015 to 2018 & 0.50 & \hl{0.625} & 0.25 & 0.375 & 0.25 & 0.125 & 0.25 & 0.125\tabularnewline
		& \textit{Interest rate changes} &  &  &  &  &  &  &  & \tabularnewline
		& 2007 to 2010 & \hl{0.50} & \hl{0.50} & 0.00 & 0.00 & 0.25 & 0.00 & \hl{0.50} & 0.00\tabularnewline
		& 2011 to 2014 & 0.125 & 0.00 & 0.50 & 0.00 & \hl{0.625} & 0.00 & \hl{0.625} & 0.00\tabularnewline
		& 2015 to 2018 & 0.375 & \hl{0.75} & 0.25 & 0.25 & 0.25 & 0.00 & 0.375 & 0.125\tabularnewline
		& \textit{Commodity index returns} &  &  &  &  &  &  &  & \tabularnewline
		& 2007 to 2010 & 0.50 & 0.375 & 0.00 & 0.375 & 0.625 & 0.00 & \hl{0.875} & 0.50\tabularnewline
		& 2011 to 2014 & 0.125 & 0.125 & 0.375 & 0.25 & 0.625 & 0.125 & \hl{0.75} & 0.00\tabularnewline
		& 2015 to 2018 & 0.50 & \hl{0.75} & 0.125 & 0.25 & 0.50 & 0.125 & 0.375 & 0.25\tabularnewline
		&  &  &  &  &  &  &  &  & \tabularnewline
		\multicolumn{2}{l}{\textit{Both tail}} &  &  &  &  &  &  &  & \tabularnewline
		& \textit{Exchange rate returns} &  &  &  &  &  &  &  & \tabularnewline
		& 2007 to 2010 & \hl{0.625} & 0.375 & 0.125 & 0.125 & 0.25 & 0.25 & 0.25 & 0.25\tabularnewline
		& 2011 to 2014 & 0.125 & 0.00 & 0.00 & 0.00 & 0.625 & 0.00 & \hl{0.875} & 0.00\tabularnewline
		& 2015 to 2018 & 0.375 & \hl{0.625} & 0.25 & 0.25 & 0.25 & 0.125 & 0.25 & 0.00\tabularnewline
		& \textit{Interest rate changes} &  &  &  &  &  &  &  & \tabularnewline
		& 2007 to 2010 & \hl{0.50} & \hl{0.50} & 0.00 & 0.00 & 0.25 & 0.00 & \hl{0.50} & 0.00\tabularnewline
		& 2011 to 2014 & 0.25 & 0.125 & 0.50 & 0.00 & \hl{0.625} & 0.00 & \hl{0.625} & 0.00\tabularnewline
		& 2015 to 2018 & 0.25 & \hl{0.75} & 0.25 & 0.25 & 0.25 & 0.00 & 0.375 & 0.125\tabularnewline
		& \textit{Commodity index returns} &  &  &  &  &  &  &  & \tabularnewline
		& 2007 to 2010 & 0.375 & 0.375 & 0.375 & 0.375 & \hl{0.625} & 0.00 & \hl{0.75} & 0.50\tabularnewline
		& 2011 to 2014 & 0.125 & 0.125 & 0.375 & 0.125 & \hl{0.625} & 0.25 & \hl{0.625} & 0.00\tabularnewline
		& 2015 to 2018 & 0.50 & \hl{0.75} & 0.125 & 0.125 & 0.50 & 0.125 & 0.375 & 0.25\tabularnewline
		\bottomrule
	\end{tabular}\hspace{0.1cm} {\small{}\medskip{}
		\setlength{\tabcolsep}{6pt}}
\end{table}
\textcolor{blue} {These average score results show that the most competitive of the proposed FQ models is for the interest rate returns. An explanation for why static models become particularly competitive for interest rates is that they have less conditional heteroscedasticity than the other data, and that most of the variation of interest rates is captured by just two components, as we have seen in \autoref{Figure:VarianceExplained}. Another possible explanation for the competitive performance of FQ forecasts, relative to those generated by EDFs  is a smoothing effect due to sampling 25,000,000 times from each predictive distribution for the FQ models, versus the available 250 data points in the calibration set for the EDF
specification.} It is also worth pointing out that our proposed FQ models outperform the static EDF approach most of the time. By the same token,  the MCS results based on proper univariate scoring rules also indicate favourable forecasting performance of both FQ specifications, matching or exceeding the accuracy of more complicated GARCH models and significantly surpassing the accuracy of copula models with EDF marginals for interest rates. The FQ models also perform competitively for the commodity indices. 

We observe that FQ and EDF models based on 250 observations almost always outperform their counterparts with 2,000 observations. This may be explained by a violation of the stationarity assumption for the data generating process over very long calibration windows. \textcolor{blue}{The EDF model performance is generally worse than that of both FQ models}. In fact, it only performs well for  1-year interest rates  and gold. This indicates that the use of latent principal component factors succeeds at reducing the amount of unimportant variation in the observed historical data and produces significantly more accurate forecasts.

  \autoref{Table:univariateMCSALSub} summarizes the MCS inclusion rates for each model over all data and then in three sub-periods that are the same for each data set: (1) March 2007 -- December 2010; (2) January 2011 -- December 2014; and (3) January 2015 -- December 2018.  As well as dividing the out-of-sample period we report results for three different weighting of CRPS as before, i.e. uniformly, centre and both-tail weighted.   The first sub-period is a little less than 4 years, because the exchange rate data  began in 1999 and we require 2000 observations to calibrate the models. \textcolor{blue}{The results in \autoref{Table:univariateMCSALSub}} demonstrate that forecasting accuracy varies strongly over time, especially for exchange rates which exhibit the most pronounced regime-specific behaviour. Most other studies in the literature evaluate models only on small samples, spanning limited time periods. Our sub-sample analysis shows that, while the proportion of MCS which include a given model does depend on the sample, the FQ models are still highly competitive, provided they are calibrated on a small sample.

\subsection{Multivariate Forecasting Accuracy Results}
\label{Section:MultivariateResults}
For the evaluation of multivariate forecasting accuracy we apply the energy score and the variogram scores with $p=0.5,1,2$. These values of $p$ were introduced by \citet{scheuerer2015} and are considered typical choices \citep{jordan2017}. Contrary to the CRPS results, the multivariate scoring rules encapsulate the accuracy for all eight marginals and their dependency into a single score which holistically quantifies the performance of the model on a given data set. 
\begin{table}[!htb]
\small
\centering 
\caption{Average values for multivariate scores.} 
\vspace{-0.3cm}
\caption*{\footnotesize {Average variogram and energy scores for different multivariate models applied to the full out-of-sample period in each of the three multivariate data sets. The variogram scores are based on $p = 0.5, 1$ and $2$ and results in each row are multiplied by  a relevant power of 10  for ease of presentation -- there is no comparison between rows. But within each row we compare the average score across the columns  and, as before, depict the lowest score in blue and use {*} and {*}{*} to 
		indicate that the model is in the 90\% and 75\% MCS, respectively.
		The bottom row summarizes the percentages each model is in the 90\%
		MCS for all the data.}
	\vspace{-0.1cm}
	\label{Table:multivariateMCS-1} }%
\begin{tabular}[t]{>{\raggedright}p{3cm}llllllll}
	\toprule 
	\multirow{1}{3cm}{Asset} & \multicolumn{2}{c}{$\mbox{GARCH}$} & \multicolumn{2}{c}{$\mbox{EDF}$} & \multicolumn{2}{c}{$\mbox{FQ-B}$} & \multicolumn{2}{l}{$\mbox{FQ-A}$}\tabularnewline
	& \multicolumn{1}{l}{CCC} & \multicolumn{1}{l}{DCC} & \multicolumn{1}{l}{$250$} & \multicolumn{1}{l}{$2000$} & \multicolumn{1}{l}{$250$} & $2000$  & $250$  & $2000$\tabularnewline
	\midrule 
	\multicolumn{4}{l}{\textit{Exchange rate returns}} &  &  &  &  & \tabularnewline
	$\mbox{VS}_{0.5}$ ($\times10^{-3}$)  & 71.37 & 67.96{*}{*} & 70.96 & 74.98 & 74.08 & 80.69 & \hl{67.67{*}{*}} & 70.50{*}\tabularnewline
	$\mbox{VS}_{1.0}$ ($\times10^{-4}$)  & 22.13 & 21.11{*}{*} & 22.20 & 22.90 & 22.92 & 24.12 & \hl{21.02{*}{*}} & 21.44{*}{*}\tabularnewline
	$\mbox{VS}_{2.0}$ ($\times10^{-7}$)  & 63.08{*}{*} & \hl{62.48{*}{*}} & 64.49 & 63.69 & 64.20 & 63.48 & 63.17{*}{*} & 62.68{*}{*}\tabularnewline
	ES ($\times10^{-3}$)  & 12.77 & \hl{12.73{*}{*}} & 12.91 & 13.03 & 12.91 & 13.10 & 12.86 & 12.98\tabularnewline
	&  &  &  &  &  &  &  & \tabularnewline
	\multicolumn{3}{l}{\textit{Interest rate changes}} &  &  &  &  &  & \tabularnewline
	$\mbox{VS}_{0.5}$  & 49.64 & \hl{45.86{*}{*}} & 64.41 & 75.50 & 66.49 & 78.22 & 66.62 & 77.85\tabularnewline
	$\mbox{VS}_{1.0}$ ($\times10^{2}$)  & 7.21 & \hl{6.41{*}{*}} & 9.17 & 10.66 & 9.11 & 10.50 & 9.15 & 10.44\tabularnewline
	$\mbox{VS}_{2.0}$ ($\times10^{4}$)  & 54.05 & 58.04 & 31.58 & 30.57 & 30.24 & 27.64 & 30.39 & \hl{27.54{*}{*}}\tabularnewline
	ES ($\times10^{-1}$)  & \hl{87.49{*}{*}} & 87.89 & 87.90 & 89.94 & 87.61{*}{*} & 89.87 & 87.59{*}{*} & 89.87\tabularnewline
	&  &  &  &  &  &  &  & \tabularnewline
	\multicolumn{4}{l}{\textit{Commodity index returns}} &  &  &  &  & \tabularnewline
	$\mbox{VS}_{0.5}$ ($\times10$)  & 22.50 & 22.66 & 20.75 & 21.15 & 20.98 & 21.51 & \hl{20.62{*}{*}} & 20.98\tabularnewline
	$\mbox{VS}_{1.0}$  & 16.58 & 16.68 & 14.80 & 15.10 & 14.89 & 15.26 & \hl{14.70{*}{*}} & 14.98\tabularnewline
	$\mbox{VS}_{2.0}$ ($\times10^{-4}$)  & 96.80 & 96.52 & 79.10 & 79.77 & 79.05 & 79.68 & \hl{78.50{*}{*}} & 79.11{*}\tabularnewline
	ES  & 33.88 & 33.89 & 33.78 & 33.93 & \hl{33.73{*}{*}} & 33.93 & 33.83 & 34.00\tabularnewline
	&  &  &  &  &  &  &  & \tabularnewline
	\textit{Summary}  & 16.7\%  & 50.0\%  & 16.7\%  & 0.0\%  & 16.7\%  & 0.0\%  & 58.3\%  & 41.7\%\tabularnewline
	\bottomrule
\end{tabular}
\end{table}

  {}\autoref{Table:multivariateMCS-1}  reports the average values obtained using different multivariate scoring rules. The results are obtained by applying each multivariate model to each asset class as a whole, \textcolor{blue}{and deriving scores from the entire out-of-sample period}.  As before, preferred models have smaller average scores. The relative accuracy of a given model depends on the scoring rule applied. The energy and variogram scores differ in their recommendations and none of the scores predominantly favours a specific model, rather the preferred model depends on the data. Overall, the FQ-A models perform the best in terms of the average scores and they are also included in more of the 90\% MCS. By comparison, there is a 50\% inclusion rate of DCC-GARCH, which is much stronger than CCC-GARCH and all the EDF models. 

The comparable performance of FQ models, even with a simple Gaussian copula, to DCC-GARCH is especially relevant since DCC-GARCH is much more computationally intensive. As pointed out in \autoref{Section:Data}, both FQ versions are at least 5 times faster and do not require additional attention to check for mis-calibrated parameters. \textcolor{blue}{Notably, FQ models again outperform EDF forecasts, despite sharing the same calibration window and the same copula.} This demonstrates that the variation reduction through our latent factor model improves the accuracy of the distribution forecast considerably.

\textcolor{blue}{In comparison with} the univariate analysis, models with longer calibration windows perform better and are now present in the superior sets. This might be because the standard errors in the correlation matrix decrease as the sample size increases. Further, the performance of DCC-GARCH is much better in the multivariate comparison than in the prior univariate one, even for exchange rate returns where CCC-GARCH was included in more superior sets than DCC-GARCH. This suggests that the time-varying conditional correlation structure is an improvement over the constant conditional correlation that requires strong assumptions which are not fulfilled for many assets.

\begin{table}[!htb]
    \centering
    \small
    \caption{Comparison of multivariate performance over time}
    \vspace{-0.3cm}
    \caption*{\footnotesize This table lists the number of times each model is included in one of the 90\% superior sets for the multivariate scores. Since we consider 4 different scoring rules, each model can be included at most 4 times.  We again use blue to highlight the most successful model in each row. Column ($\ast$) uses the entire out-of-sample periods while columns (1), (2) and (3) are restricted to the sub-periods March 2007 -- December 2010, January 2010 -- December 2014 and January 2015 -- December 2018 respectively.}
    \label{Table:multivariatePerformancePeriods}
    \begin{tabular}[t]{lllllllll}
    		\toprule 
    	& \multicolumn{2}{c}{\hspace{-0.05cm}$\mbox{GARCH}$} & \multicolumn{2}{c}{\hspace{-0.05cm}$\mbox{EDF}$} & \multicolumn{2}{c}{\hspace{0cm}$\mbox{FQ-B}$} & \multicolumn{2}{l}{$\mbox{FQ-A}$}\tabularnewline
    	& \multicolumn{1}{l}{CCC} & \multicolumn{1}{c}{DCC} & \multicolumn{1}{l}{$250$} & \multirow{1}{*}{$2000$} & \multicolumn{1}{c}{$250$} & $2000$  & $250$  & $2000$\tabularnewline
    	\midrule 
    	\multicolumn{1}{l}{\textit{Exchange rate returns}} &  &  &  &  &  &  &  & \tabularnewline
    	2007 to 2010  & 1 & \hl{4} & 0 & 3 & 1 & 2 & 1 & 3\tabularnewline
    	2011 to 2014  & 1 & \hl{2} & 0 & 1 & 1 & 0 & 1 & 1\tabularnewline
    	2015 to 2018  & 1 & 0 & 0 & 3 & 1 & 3 & \hl{4} & 3\tabularnewline
    	\textit{Interest rate changes}  &  &  &  &  &  &  &  & \tabularnewline
    	2007 to 2010  & 0 & 2 & 0 & 0 & 0 & 0 & \hl{3} & 0\tabularnewline
    	2011 to 2014  & 1 & \hl{2} & 0 & 0 & \hl{2} & 0 & 1 & 0\tabularnewline
    	2015 to 2018  & 0 & 0 & 1 & 0 & 1 & 0 & \hl{3} & 0\tabularnewline
    	\textit{Commodity index returns}  &  &  &  &  &  &  &  & \tabularnewline
    	2007 to 2010  & 0 & 2 & 1 & 0 & 2 & 0 & \hl{4} & 2\tabularnewline
    	2011 to 2014  & 1 & \hl{3} & 0 & 0 & 2 & 0 & 2 & 0\tabularnewline
    	2015 to 2018  & 0 & 0 & \hl{4} & 1 & 3 & 1 & 3 & \hl{4}\tabularnewline
    	\bottomrule
    \end{tabular}\hspace{0.1cm} {\small{}\medskip{}
    	\setlength{\tabcolsep}{6pt}}
\end{table}

\autoref{Table:multivariatePerformancePeriods} investigates  the robustness of the results in \autoref{Table:multivariateMCS-1} by separating the out-of-sample period into 3 sub-periods, as before. Here we only report how many of the four multivariate scoring rules include each particular model in the superior set. For instance, the number 3 for the CCC-GARCH applied to exchange rates for the  sub-sample 2007 -- 2010 indicates that 3 out of 4 scoring rules keep this model in the MCS, when scores are derived only from this sub-sample.  According to this criteria, the ranking depends on  the sub-sample and in most samples the highest rank is accorded to DCC-GARCH or FQ-A$_{250}$.
\section{Summary and Conclusions} 
\label{Section:Conclusions}
This paper contributes to the empirical analysis of proper multivariate scoring rules and introduces a new class of Factor Quantile (FQ) models. The FQ models are flexible and semi-parametric, and they are employed here to generate multivariate distribution functions where marginals are derived from interpolations on quantiles estimated via factor-model regressions and their dependence is selected by choosing a parametric conditional copula. It is not a dynamic model but we demonstrate that its forecasts, based on the idea that \textcolor{blue}{the joint distribution} of the variables can be well approximated by their joint historical distribution, are at least as accurate as constant and dynamic conditional correlation models with Student-t asymmetric E-GARCH(1,1) marginals. Moreover, the FQ models have several advantages over multivariate GARCH models. These should make them attractive to banks and asset managers -- or any other players involved in portfolio optimisation and multi-asset pricing -- who aim to model and/or forecast large multivariate distributions of financial asset returns.

The class of FQ models is very flexible: they can be built on any factor model and they can use any conditional copula. We have illustrated an application of the FQ model to bivariate stock returns using the asymmetric CAPM factor model with a Gumbel copula. However, in larger-dimensional systems we strongly advocate the use of latent principal component factors, for which we have developed two alternative versions. One of them is very simple to implement and the other requires the use of a bagging algorithm proposed by \citet{breiman1996}. 

Our extensive empirical study forecasting exchange rates, interest rates and commodity futures  is the first substantial study of multivariate distribution forecasting for financial asset returns. We assess the accuracy of forecasts using the MCS of \citet{hansen2011} derived from the (strictly) proper energy score \citep{szekely2003}, the variogram score \citep{scheuerer2015} and the weighted CRPS introduced by \citet{gneiting2007b}.  This way, we highlight how both the scores and the superior model sets depend on the asset class and the sample. These conclusions accord with  \citet{giacomini2006}, \citet{machete2013} and \citet{elliott2016}, all of whom emphasise that there is no single superior approach: the best model depends on the statistical properties of the data and the economic properties of the variable being predicted. 

\textcolor{blue}{The best univariate model for each data  set depends on the weights used in the CRPS, but overall, these scores favour GARCH models for exchange rates and commodity indices, and FQ models for interest rates. However, the results are also sample specific. For instance, in the period 2011 to 2014 the FQ-A model was best for each data set, according to the tailed-weighted CRPS, but  in the period 2015 to 2018 the GARCH-DCC model was best for each data set, according to the same scoring rule. The multivariate results which are based on energy scores and variogram scores with  different parameters also depend on the scoring rule employed. For exchange rates, the best models are: FQ-A(250) according to the variogram scores with $p=0.5$ and $p=1$; and GARCH-DCC according to the variogram score with $p=2$ and the energy score. For interest rates, the best models are: GARCH-DCC, according to the variogram scores with $p=0.5$ and $p=1$; FQ-A(2000) according to the variogram score with $p=2$; and GARCH-CCC according to the energy score.  For commodity indices, the best models are: FQ-A(250) according to the variogram scores with $p=0.5, 1$ and $2$; and the EDF(250) according to the energy score. The reason why different rankings are obtained from different scoring rules applied to diverse data sets may be an interesting subject for further research. }

\textcolor{blue}{Overall, our empirical results suggest that the FQ-B models perform slightly worse
	than the FQ-A models. Since the latter are conceptually simpler and much easier
	to implement, these may be the preferred choice in practical applications. The bagging algorithm used in the FA-B model is computationally complex, and it should only be applied in practice  if it provides a clear improvement on other FQ models.}

We conclude that the forecasting performance of latent factor FQ models \textcolor{blue}{generally} exceeds the static model that is standard in the industry, i.e. historical simulation (the variant represented in this paper uses a Gaussian copula with EDF marginals). The forecasts generated by these FQ models also match or exceed the accuracy of standard  dynamic forecasting models, represented here with multivariate GARCH. However -- and even though we have not taken the most advanced models in the class -- the multivariate GARCH models still take over five times longer to optimize, require very large calibration samples and  exhibit difficulties with parameter convergence even in eight dimensions. By contrast, FQ models scale naturally to high-dimensional systems while also retaining very fast calibration times. They are much easier to fully automate than GARCH models and could therefore be very  attractive to hedge funds and other high-frequency traders in the industry, who commonly employ algorithms to re-balance portfolios every day.



\newpage
\singlespacing


\footnotesize
\bibliography{BibTeX}
\bibliographystyle{agsm}
\clearpage

\end{document}